\documentclass[doc, 12pt]{apa7} 

\usepackage[american]{babel}
\usepackage{csquotes}
\usepackage[style=apa, sortcites=true, sorting=nyt, backend=biber]{biblatex}
\DeclareLanguageMapping{american}{american-apa}
\usepackage{mathptmx} 

\usepackage{amsmath}
\usepackage{bm}
\usepackage{txfonts}
\usepackage{url}
\usepackage{tikz} 
\usetikzlibrary{positioning, calc}
\usetikzlibrary{positioning, arrows.meta, fit}
\usepackage{xcolor} 
\usepackage{longtable}
\usepackage{array}
\usepackage{hyperref}
\usepackage{nameref}
\usepackage{float} 
\usepackage{comment}

\title{Reconciling Latent Variables and Networks: Exploring and extending the Psychometric-Toolbox}
\shorttitle{prePrint - 13.11.26 - This paper has not been peer reviewed.} 
\authorsnames[1,2,1,1]{Kevin Kistermann, Maarten Marsman, Vivato V. Andriamiarana, Augustin Kelava}
\authorsaffiliations{
{University of Tübingen (Methods Center)},
{University of Amsterdam}
}
\authornote{\addORCIDlink{Kevin Kistermann}{0009-0006-2822-1245}: Conceptualization, Visualization, Writing—original draft. \addORCIDlink{Maarten Marsman}{0000-0001-5309-7502}: Writing—review and editing. \addORCIDlink{Vivato V. Andriamiarana}{0009-0000-7862-6319}: Writing—review and editing. \addORCIDlink{Augustin Kelava}{0000-0001-6053-0415}: Conceptualization, Funding acquisition, Project administration, Supervision, Writing—review and editing. The authors report there are no competing interests to declare and generative AI was not used in their research or preparation of this manuscript. Correspondence concerning this article should be addressed to Kevin Kistermann, Methods Center Institute, Eberhard-Karls Universität Tübingen. E-Mail: kevin.kistermann@mz.uni-tuebingen.de\\
An earlier version of this manuscript has been made available as a preprint \parencite{Kistermann.2026.prePrint}.
\paragraph{Code \& Data} The code is available at \url{https://github.com/kekistermann/Exploring-the-PsychometricToolbox} (\href{https://doi.org/10.5281/zenodo.24187806}{DOI: 10.5281/zenodo.24187806}), and the data at \url{https://osf.io/jmz2n/overview}.
\paragraph{Funding} Funding for this research was provided by the Deutsche Forschungsgemeinschaft (Project number 528244660).}

\abstract{Since the introduction of network psychometrics, several connections to statistical models in “classical” psychometrics (i.e., Factor Analysis, Structural Equation Modeling, Item Response Theory) as well as to approaches from other research fields have been established. This paper reviews, synthesizes, and extends these developments, presenting them in an accessible visual format. This perspective opens up promising opportunities to extend the psychometric-toolbox by incorporating and learning from statistical methodologies developed in other areas of research, which often address similar or even identical problems. Highlighting these methodological commonalities may also foster collaboration across research fields that have traditionally remained largely independent. Moreover, awareness of these connections may render methodological development more systematic and goal-directed and may enable a meaningful division of labor, for example between the development of statistical methodology and its practical implementation for empirical research through software tools. Finally, these methodological advances provide new opportunities for empirical research and may contribute to a reconciliation with longstanding conceptual issues concerning psychometric constructs and, more broadly, psychological phenomena.}


\begin{document}
\tableofcontents
\newpage
\maketitle
In recent years, statistical methodologies associated with "network psychometrics" or "the network approach" \parencite{Borsboom.2021.NetworkPsychometrics, Borsboom.2022.NetworkPsychometrics, Isvoranu.2022.NetworkPsychometrics} have extended the psychometric-toolbox, most notably in form of the \textit{Gaussian Graphical Model} (GGM; \cite{Epskamp.2018.GGM-GVAR, Altenbuchinger.2020.GGM, Shutta.2022.GGM, Uhler.2017.GGM}) and \textit{Ising Model} (IM; \cite{Cipra.1987.Ising, Finnemann.2021.Ising, Macy.2024.Ising}). But rather than replacing traditional latent variable approaches such as \textit{Structural Equation Modeling} (SEM; \cite{Bollen.1989.SEM, Tarka.2018.SEM, Westland.2019.SEM, Hoyle.2023.SEM}) and \textit{Item Response Theory} (IRT; \cite{Birnbaum.1968, Ayala.2009.IRT, vanderLinden.2018.IRT}), these models should ideally be viewed as complementary, particularly because it has been shown that from a purely statistical perspective they are closely related and, in some cases, even equivalent \parencite{Golino.2017.EGA, Epskamp.2017.LNM-RNM, Marsman.2018.IRT-Ising, Wang.2021, vanBork.2021, Waldorp.2022}. In addition to that, at the conceptual level, the critical necessity for more individual-level (idiographic) research in psychology has been stressed \parencite{Molenaar.2004.IdiographicPsy, Piccirillo.2019.IdiographicPsy, Chow.2024.IdiographicPsy, Borsboom.2024.IdiographicPsy,}, which necessarily implies a dynamic understanding of constructs and therefore the need for time-series data. And with the increasing feasibility of Ambulatory Assessment (AA; \cite{Trull.2014.AA}) protocols\footnote{also Experience Sampling Methods (ESMs), or Ecological Momentary Assessments (EMAs)}, a method for collecting Intensive Longitudinal Data (ILD; \cite{Hamaker.2025.ILD}), the need for developing appropriate statistical models to analyze such high-dimensional data is critical. Modeling frameworks for doing so have been developed in psychometrics mostly in the form of dynamic latent variable models \parencite{Asparouhov.2018.DSEM, Asparouhov.2017.DLCA, Kelava.2019.NDLC-SEM}, but the network approach has been extended to the time-series domain as well \parencite{Epskamp.2018.GGM-GVAR, Epskamp.2020.panel/ts-lvgvar}. Here again, similarities in model formulation have been observed \parencite{Kelava.2019.NDLC-SEM, Epskamp.2020.panel/ts-lvgvar}. This prompted the review and synthesis presented below, which features an accessibly visualization\footnote{Inspired by \textcite{Leemis.2008}; \url{https://www.math.wm.edu/~leemis/chart/UDR/UDR.html}} in Figure \ref{fig:ModelDiagram}, \ref{fig:ModelDiagram_2}, \ref{fig:ModelDiagram_3}, \ref{fig:ModelDiagram_4} (and 11 in Supplementary Materials). Each statistical model is represented as a colored square, with the colors roughly representing different areas of research where the model originated and the relations are labeled when possible with its key characteristic. The full diagram, including models omitted from this paper due to space limitations, together with a more detailed description of the relationships, is available at: \url{https://github.com/kekistermann/Mapping-the-PsychometricToolbox}.

The paper is primarily divided into three sections. In \textit{\nameref{1Section}}, we start by going over some of the previously established relations in the psychometric literature, this includes the IM and certain IRT models \parencite{Marsman.2018.IRT-Ising}, the GGM and \textit{Factor Analysis} (FA; \cite{Bartholomew.2011.GLLVM}) in the cross-sectional \parencite{Golino.2017.EGA, Epskamp.2017.LNM-RNM} as well as time-series domain \parencite{Epskamp.2018.GGM-GVAR}, and more recently Random Graph models \parencite{Marsman.2023.IdiographicIsing, Marsman.2023b.IdiographicIsing}. With the primary aim of providing an accessible review and synthesis while also serving as a foundation for future exploration of additional relationships. This should highlight how seemingly distinct statistical methodologies from independent research fields, scattered throughout the literature, are in fact closely related (e.g., \cite{MaydeuOlivares.2026.psychometrics---econometrics}). By bringing these connections together, it aims to foster interdisciplinary collaboration and open up valuable opportunities for future research. For applied researchers, it simplifies this complex landscape and may facilitate the selection of appropriate tools for more robust theoretical interpretations. Furthermore, as data structures become increasingly complex, close collaboration between applied researchers and statistical experts (e.g., methodologists and data scientists) becomes essential \parencite{Grimm.2026}. Lastly, introductory tutorials for three models (multilevel VAR, GIMME, and Recurrence Plot) are provided, analyzing a publicly available psychometric ILD dataset \parencite[from][]{Rowland.2020b.Dataset}. And in the supplementary materials many additional relations are pointed out, which could extend statistical methodology in psychometrics in novel ways.

In \textit{\nameref{2Section}} we briefly highlight that Network Science \parencite{Sweet.2025}, Graph Theory \parencite{Das.2023.GraphDataScience, Diestel.2025.GraphTheory}, and network psychometrics represent fairly independent research fields, the latter of which is---at least currently---primarily concerned with the estimation of Probabilistic Graphical Models \parencite{Maasch.2025.PGM} on psychometric datasets. But some formal connections already exist \parencite{Farasat.2015, Marsman.2023.IdiographicIsing}, and some basic psychometric concepts can be re-interpreted in the network perspective \parencite{Borsboom.2023}. We also refer to some statistical tools for modeling more than pairwise interactions as some form of covariance, which is not always appropriate \parencite{Slipetz.2024.dcorr}, but currently mostly the default. Many more statistical measures are, in principle, available here \parencite[see][]{Cliff.2023.SPI}. 

And \textit{\nameref{3Section}} tries to contextualize that network psychometrics not only provides us with new statistical tools, but it is perhaps most productively framed as enabling a conceptual rethinking of psychometric constructs. In that light it is important to keep in mind the distinction between statistical models and theories of psychological phenomena \parencite{vanBork.2021, vanDongen.2024}. And that the development of formal theories in the form of computational models is a recent development that might be helpful in bridging that gap \parencite{Guest.2021, Woensdregt.2024, Robinaugh.2024.PanicDisorder, Evers.2026.PsyTheory}. Networks as a conceptual framework also provide a natural connection between idiographic psychological processes, complex dynamical systems, and Graph Theory as the appropriate mathematical formalism \parencite{Marsman.2023.IdiographicIsing, vanderMaas.2024,Waldorp.2025.Perturbation-Graphs, Vogel.2025.PCN}. This potentially represents access to a plethora of methods that could be utilized in psychometrics more rigorously. And perhaps most interestingly, this connects naturally to a broad discussion in the philosophy of science about what constitutes appropriate and useful explanations \parencite{Kostic.2025, Huneman.2025}.

\section{Exploring and extending the Psychometric-Toolbox} \label{1Section}
\begin{figure*}[!htbp]
    \centering
    \includegraphics[scale=0.3]{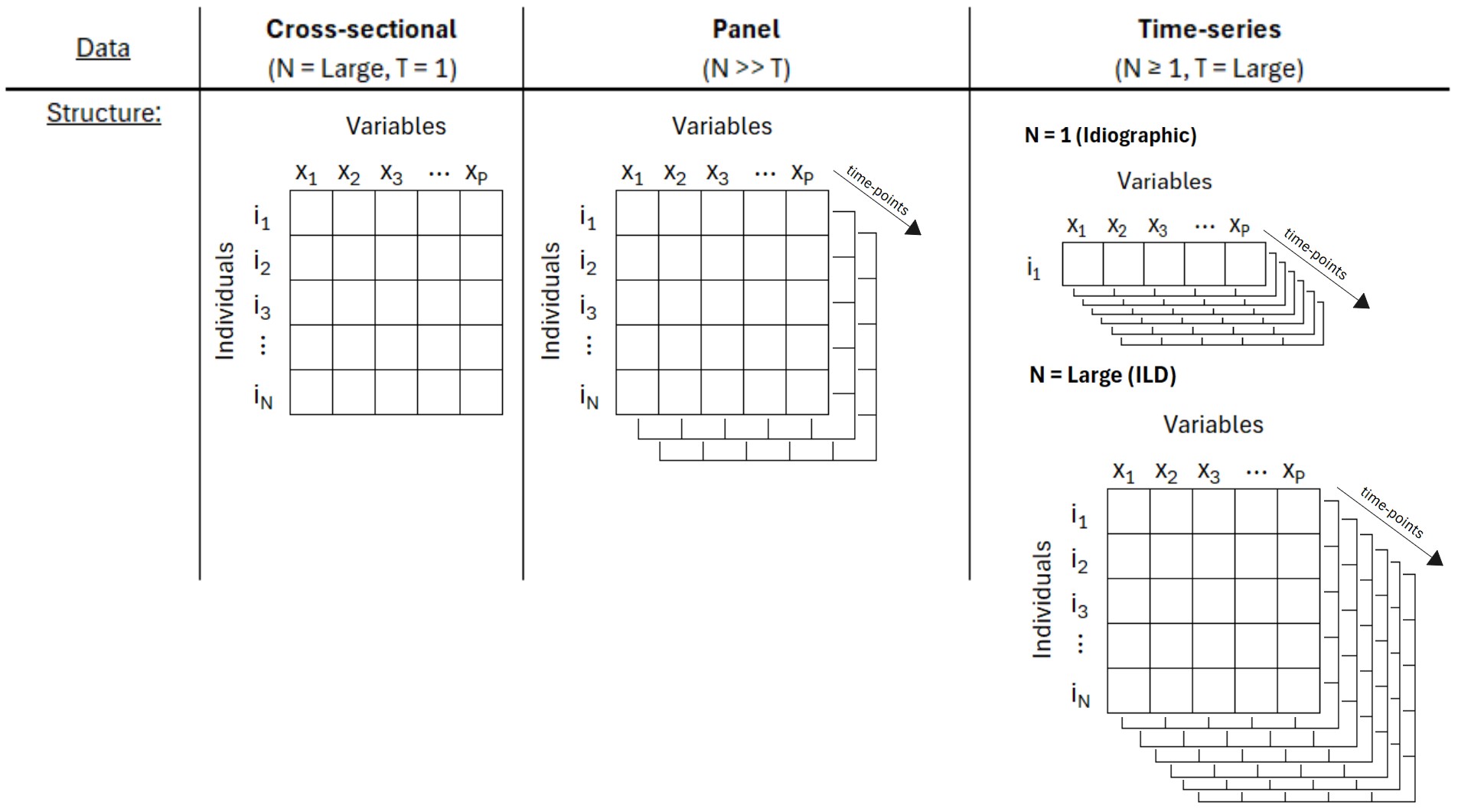}
    \caption{A useful heuristic for differentiating between common basic data structures in empirical research was introduced by \cite{Cattell.1952.DataCube} and termed 'Data-Box/Cube' in \parencite{Molenaar.2012, Wardenaar.2013.DataCube, Ram.2015.DataCube, Robinaugh.2020.Psychopathology, Isvoranu.2022.NetworkPsychometrics, Borsboom.2024.IdiographicPsy}. It has three dimensions [individuals ($N$) × variables ($P$) × time-points ($T$)] which classifies the four common data structures illustrated here. \parencite[Figure modified from][]{Borsboom.2021.NetworkPsychometrics}}
    \label{fig:Cartell-DataBox}
\end{figure*}

Statistical models can be broadly classified along the lines of three dimensions, with $N$ indicating the number of individuals, $P$ the number of variables, and $T$ the number of time points (measurement occasions), which determines their applicability to the four common data structures visualized in Figure \ref{fig:Cartell-DataBox}. The variables $P$ can then be further subdivided into different types (binary, categorical, continuous).

\subsection{Initial Relations: Item Response Theory and the Ising Model (Figure \ref{fig:ModelDiagram})}
We start by considering a cross-sectional dataset $\bm{X} = [\bm{X}_1, ... , \bm{X}_P]$, with $P$ binary variables collected from many individuals ($N$=Large) at one time-point ($T$=1). In IRT these binary variables commonly represent correct ("1") or incorrect ("0") responses to questions (usually referred to as items) in a psychometric test. It is assumed that the item responses are explained by a set of $M$ $(M \le P)$ latent variables $\bm{\Theta} = [\bm{\theta}_1, ..., \theta_M]$ that usually represent some form of ability in the educational testing context where IRT originated and is primarily applied. One of the simplest models in IRT for the response vector $\bm{x} = [x_1, ..., x_P]$ of one individual, is the \textit{Rasch model} (\cite{Rasch.1960.IRT}; or 1-Parameter Logistic (1PL) model) in Equation (\ref{eq_1}).
\begin{equation}
    p(\bm{x} | \theta)_{1PL} = \prod_{i} \frac{\exp{(x_i \alpha (\theta - \delta_i))}}{Z_{1PL}} \tag{1} \label{eq_1}
\end{equation}
With $\delta_i$ being the difficulty parameter of item $i$ and $\alpha$ the discriminating parameter (effectiveness of the item to differentiate between different ability levels) for all items. When $\alpha$ is allowed to vary between items we get the \textit{2-Parameter Logistic} (2PL) model. With more than one latent variable $(M > 1)$,  we get the \textit{Multidimensional 2PLM} (MD-2PLM; \cite{Reckase.2009.MD-2PLM}) model in Equation (\ref{eq_1a}).
\begin{equation}
    p(\bm{x} | \bm{\theta})_{MD-2PLM} = \prod_{i} \frac{\exp{(x_i (\bm{\alpha}_i^\top \bm{\theta} - \delta_i))}}{Z_{MD-2PLM}} \tag{1a} \label{eq_1a}
\end{equation}
Many models in the IRT framework \parencite{Chen.2025.IRT} are a variant of the same idea, namely, it defines a functional structure which specifies how people’s position on a latent variable, usually a monotonically increasing function called item characteristic curve (ICC), "causes" a probability distribution over the item responses \parencite{Mellenbergh.1994}. And they usually rely on some form of conditional independence, which means that the item responses are independent given the latent variable, and can therefore be understood as specifying a common-cause model \parencite{Borsboom.2023}.
\paragraph{Network Psychometrics}
In the emerging research paradigm of network psychometrics on the other hand, it is assumed that the item responses are neither caused (\textit{reflective/common cause framework}) by, or being the cause (\textit{formative/common effect framework}) of latent variables, but rather that they represent variables as mutually affecting each other (\textit{reciprocal effect framework}) in a causally related network \parencite{Kruis.2016.IRT-Ising}. The network approach has been applied to explain many psychometric constructs, for example in intelligence research \parencite{vanderMaas.2006.Intelligence, vanderMaas.2017.Intelligence, Savi.2019.Intelligence, McGrew.2023.Intelligence}, personality research \parencite{Costantini.2015.Personality, Christensen.2020.Personality, Lunansky.2020.Personality-Psychopathology}, and probably most extensively in psychopathology\footnote{e.g., Depression \parencite{Cramer.2016.Depression}, Social Anxiety \parencite{Noda.2025.SocialAnxiety}, Alcohol Use Disorder \parencite{Oostendorp.2025.AlcoholUseDisorder}, Post-Traumatic Stress Disorder \parencite{Robinaugh.2025.PTSD}, and affect dynamics in general \parencite{Loossens.2020.Affect}} research \parencite{Borsboom.2013.Psychopathology, Bringmann.2018.Psychopathology, Robinaugh.2020.Psychopathology, Briganti.2024.Psychopathology, Kashihara.2025.Psychopathology}.

\paragraph{Ising Model}
The first statistical model associated with this approach was the IM from statistical physics in Equation (\ref{eq_2}). Here the binary variables in the dataset $\bm{X}$ represent the orientation of particles, which are either in the state\footnote{The Ising model can have $\{-1, 1\}$ or $\{0, 1\}$ coding, both can be transformed to represent the other, but they have different parameter estimates and imply different dynamics \parencite[see][for details]{Haslbeck.2021.Ising}. For example, the $\{1, -1\}$ parameterization may be more plausible for variables that are not qualitatively different, but rather opposing each other in some way (e.g., agreeing or disagreeing with a statement). The $\{0, 1\}$ encoding could be more appropriate if the variable states are qualitatively different (e.g., presence or absence of psychiatric symptoms).} of spin-up (encoded as "+1") or spin-down ("-1") and arranged on a square lattice graph.
\begin{equation}
    p(\bm{x})_{Ising} = \frac{1}{Z_{Ising}} \exp{\left(  \sum_{i} x_i \mu_i + \sum_{<i,j>} x_i x_j \sigma_{ij} \right)} \tag{2} \label{eq_2}
\end{equation}
With $\mu_i$ as main effect or threshold for changes in variable state, for positive main effects ($\mu_i>0$) variables are more likely to have a positive value ($X_i=+1$) and with negative main effects ($\mu_i<0$) tend to have a negative value ($X_i=-1$). Pairwise interactions between variables are parametrized with $\sigma_{i j}$, with $X_i$ and $X_j$ being more likely in the same state with a positive interaction effect ($\sigma_{i j}>0$), but in different states if it is negative $\left(\sigma_{i j}<0\right)$. $Z_{Ising}$ is a normalizing constant and $\sum_{<i,j>}$ the sum of all node pairs $(i,j)$ that are direct neighbors on the lattice graph (see Figure \ref{fig:IsingLatticeGraph}).

\begin{figure}[!htbp]
    \centering
    \begin{tikzpicture}[>=stealth, node distance=2cm]
    \node[rectangle, draw, minimum size=1.2cm] (A) {$X_1$};
    \node[rectangle, draw, below=of A, minimum size=1.2cm] (B) {$X_2$};
    \node[rectangle, draw, right=of B, minimum size=1.2cm] (C) {$X_3$};
    \node[rectangle, draw, right=of A, minimum size=1.2cm] (D) {$X_4$};
    \draw[->] (A) ++(-1,1) node[above] {$\mu_1$} -- (A);
    \draw[->] (B) ++(-1,1) node[above] {$\mu_2$} -- (B);
    \draw[->] (C) ++(-1,1) node[above] {$\mu_3$} -- (C);
    \draw[->] (D) ++(-1,1) node[above] {$\mu_4$} -- (D);
    \draw[-] (A) -- node[midway, right] {$\sigma_{12}$} (B);
    \draw[-] (B) -- node[midway, below] {$\sigma_{23}$} (C);
    \draw[-] (A) -- node[midway, below] {$\sigma_{14}$} (D);
    \draw[-] (D) -- node[midway, right] {$\sigma_{34}$} (C);
    \fill[black] ($(A)+(0,0.8)$) circle (1pt);
    \fill[black] ($(A)+(0,1)$) circle (1pt);
    \fill[black] ($(A)+(0,1.2)$) circle (1pt);

    \fill[black] ($(D)+(0,0.8)$) circle (1pt);
    \fill[black] ($(D)+(0,1)$) circle (1pt);
    \fill[black] ($(D)+(0,1.2)$) circle (1pt);

    \fill[black] ($(D)+(0.8,0)$) circle (1pt);
    \fill[black] ($(D)+(1,0)$) circle (1pt);
    \fill[black] ($(D)+(1.2,0)$) circle (1pt);

    \fill[black] ($(C)+(0.8,0)$) circle (1pt);
    \fill[black] ($(C)+(1,0)$) circle (1pt);
    \fill[black] ($(C)+(1.2,0)$) circle (1pt);
    \end{tikzpicture}
    \caption{The square lattice graph of the IM illustrated. With $X_p$ being binary random variables, pairwise interactions between variables are parametrized with $\sigma_{i j}$, and $\mu_i$ as main effect or threshold for changes in variable state.}
    \label{fig:IsingLatticeGraph}
\end{figure}

\paragraph{IRT --- IM}
Formal equivalences between the IM and certain IRT models have been established in \textcite{Marsman.2015.IRT-Ising, Kruis.2016.IRT-Ising, Marsman.2018.IRT-Ising,  Epskamp.2018.IRT-Ising}\footnote{Extending IRT models to probabilistic network models motivated by the need to account for violations of the local independence assumption, has been proposed independently some time ago \parencite[see][]{Ueno.2002.IRT-Networks}.}. To see how this is the case let's start by rewriting Equation (\ref{eq_2}) in matrix form:
\begin{equation}
    \exp{( \bm{x}^\top  \bm{\mu} +  \frac{1}{2} \bm{x}^\top \bm{\Sigma} \bm{x})} \tag{2a} \label{eq_2a}
\end{equation}
The $P \times P$ weight (or connectivity) matrix $\bm{\Sigma}$ encodes the pairwise interactions between variables. If all off-diagonal elements in $\bm{\Sigma}$ are equal, the IM reduces to the Curie-Weiss model which is equivalent to the extended-Rasch model (E-RM) and marginal-Rasch model (M-RM) \parencite[see][]{Marsman.2018.IRT-Ising}. When $\bm{\Sigma}$ is chosen to be positive (semi)definite, the following non-negative eigenvalue decomposition is possible:
\[
    \bm{\Sigma} + c \bm{I} = \bm{Q \Lambda Q^\top}
\]

The exponential in Equation (\ref{eq_2a}) can then be written:
\begin{equation}
    \exp{\left(\sum_{i} x_i \mu_i + \sum_{r} \frac{1}{2} \lambda_r \left(\sum_{i} q_{ir} x_i \right)^2\right)} \tag{2b} \label{eq_2b}
\end{equation}
With $\lambda_r$ as the r\textsuperscript{th} eigenvalue in $\bm{\Lambda} = [\lambda_1, ..., \lambda_R]$, and $q_{ir}$ as the i\textsuperscript{th}-row and r\textsuperscript{th}-column in the $P \times P$ eigenvector matrix $\bm{Q}$. With Equation (\ref{eq_2b}) a latent-variable\footnote{Which in a network (graph) can be understood as cliques or clusters (subset of vertices), see \textcite{Golino.2017.EGA}. And despite "the fact that these conceptual models differ markedly in their interpretation as to what causes covariances in observables, it has been previously noted that the associated statistical models are closely related. For example, \textcite{Cox.2002} showed that there is an approximate relation between the Ising model, or quadratic exponential distribution \parencite{Cox.1972}, and the IRT model of \textcite{Rasch.1960.IRT}. Moreover, \textcite[p. 82]{Molenaar.2003} specifically suggested that there exists a formal connection between the Ising model and the IRT models of Rasch and \textcite{Birnbaum.1968}. This formal connection between the Ising model and IRT models was recently established [...]" \parencite{Marsman.2018.IRT-Ising}.} representation of the IM can be obtained (see \cite{Marsman.2015.IRT-Ising, Kruis.2016.IRT-Ising} for the detailed derivation), with as many latent dimensions as non-zero eigenvalues. When we merge the sums over $i$, denote $\lambda_r q_{ir}$ as $\alpha_{ir}$ (i\textsuperscript{th}-row and r\textsuperscript{th}-column in the $P \times P$ matrix $\bm{A}$), and rewrite the sum over $r$ as the product of $\bm{\alpha}_i^\top$ (i\textsuperscript{th}-row of $\bm{A}$) with $\bm{\theta}$, we get the following:
\begin{equation}
    p(\bm{x} | \bm{\theta})_{Ising} = \prod_{i} \frac{\exp{(x_i (\mu_i} + \bm{\alpha}_i^\top \bm{\theta})}{Z_{Ising}} \tag{2c} \label{eq_2c}
\end{equation}
Which is equivalent to the 2PL-MIRT in Equation (\ref{eq_1a}), and with only one non-zero eigenvalue equivalent to the 1PL model in Equation (\ref{eq_1}). The equivalence can be characterized in summary by the following: The IM represents relations between binary random variables with a positive semi-definite weight matrix, all eigenvalues of the eigenvalue decomposition of such a matrix are non-negative and can be reformulated as a MIRT model by invoking Kac’s Gaussian identity \parencite{Kac.1968}, which facilitates the derivation of a posterior distribution for the latent variables that has as many common factors as the number of positive eigenvalues, while the corresponding discrimination parameters are determined as functions of the associated eigenvector values \parencite{vanBork.2021}.

This connection has since then been further extended, for example, \textcite{Marsman.2025a, Marsman.2025b} formulated with Ordinal MRF a network model for ordinal variables based on the generalized partial credit model \parencite{Muraki.1990}, building on earlier work by \textcite{Anderson.2000}. And on the other hand, \textcite{Gilbert.2025} applied an IRT-based framework to model causal treatment effects on network states. Together, these developments suggest that the relationship between network models and latent variable models continues to yield new insights.

\begin{figure*}[!htbp]
\centering
\begin{tikzpicture}[
    font=\small,
    model/.style={
        rectangle,
        thick,
        minimum width = 0.1cm,
        minimum height = 0.1cm,
        align = center,
        inner sep=1pt
    },
    binary-categorical/.style={
        draw,
        rectangle,
        dashed, ultra thick,
        minimum width = 0.1cm,
        minimum height = 0.1cm,
        align = center,
        inner sep=1pt
    },    
    physics_cross/.style={model, fill=blue!10},
    physics_time/.style={model, fill=blue!45},
    ML_cross/.style={model, fill=cyan!10},
    ML_time/.style={model, fill=cyan!45},
    Graph_cross/.style={model, fill=magenta!15},
    Graph_time/.style={model, fill=magenta!45},
    psychometrics_cross/.style={model, fill=green!10},
    psychometrics_time/.style={model, fill=green!45},
    network_cross/.style={model, fill=teal!10},
    network_time/.style={model, fill=teal!45},
    econometrics/.style={model, fill=orange!45},
    relation/.style={-{Latex[length=2mm]}, thick}
]
\node[physics_cross] (IM) at (-8,6) {
\textbf{Ising Model (IM)}\\[1pt]
$p(\bm{x}) \propto \exp\!\left( \sum_i^{n} \mu_i x_i + \sum_{<i,j>} \sigma_{ij} x_i x_j\right)$
};
\node[physics_cross] (CW) at (-1,7) {
\textbf{Curie-Weiss}
};
\node[physics_cross] (BC) at (-8,9.2) {
\textbf{Blume-Capel}\\[1pt]
$p(\bm{x}) \propto \text{IM} + \sum_i^{n} \alpha^2 x_i^2$
};
\node[physics_cross] (Potts) at (-10,8) {
\textbf{Potts}\\[1pt]
$x_i \in \{1, 2, 3, .., h\}$
};
\node[psychometrics_cross] (E-RM) at (1,3) {
\textbf{E-RM}\\[1pt]
$p(\bm{x}) \propto \prod_{i=1}^p \beta_i^{x_i} \lambda_{x_+}$
};
\node[psychometrics_cross] (M-RM) at (-3,5) {
\textbf{M-RM}
};
\node[psychometrics_cross] (MD-2PLM) at (-4.9,4) {
\textbf{MD-2PLM}
};
\node[ML_cross] (logistic) at (-4,8) {
\textbf{Logistic Regression}
};
\node[ML_cross] (GLM) at (1,8) {
\textbf{GLM}\\[1pt]
$g(\bm{\eta})=\bm{X}\bm{\beta}$
};
\node[ML_cross] (MRF) at (-6,2.7) {
\textbf{Markov Random Field (MRF)}\\[1pt]
$p(\bm{x}) \propto \prod_c \psi_c(\bm{x}_c)$
};
\node[network_cross] (OMRF) at (-10,4.2) {
\textbf{Ordinal MRF}
};

\draw[relation] (IM.east) -- (CW.west) node[fill=white, midway] {$\sigma_{i,j} = \sigma$};
\draw[relation] (BC) -- (IM) node[fill=white, near start] {$\alpha = 0$};
\draw[relation] (Potts) -- (IM) node[fill=white, midway] {$h=2$};
\draw[relation] (GLM.west) -- (logistic.east) node[fill=white, midway, sloped, font=\small] {logit link};
\draw[relation] (IM) -- (MD-2PLM.north) node[fill=white, midway, font=\small]{Kac’s integral};
\draw[] (E-RM) -- (CW) node[fill=white, midway] {$\begin{aligned}
    \log \beta_i &= \mu_i \\
    \log \lambda_{x_{+}} &= \sigma x_{+}^2
    \end{aligned}$};
\draw[relation] (M-RM) -- (CW.south) node[fill=white, midway] {$f(\theta) = g(\theta)$};
\draw[relation] (E-RM.west) -- (M-RM) node[fill=white, midway, font=\small] {moment seq. $\lambda$};
\draw[relation, dashed] (MD-2PLM) -- (M-RM);
\draw[relation] (IM) -- (logistic) node[fill=white, midway] {$X_i \mid \bm{X}^{(i)}$};
\draw[relation] (MRF) -- (IM.south) node[fill=white, pos=0.4, font=\small] {Binary};
\draw[relation] (MRF.west) -- (OMRF.south) node[fill=white, midway, font=\small] {Ordinal};
\draw[relation] (OMRF) -- (IM) node[fill=white, midway, font=\small] {$m=1$};

\end{tikzpicture}
\caption{Extended Rasch Model (E-RM), Marginal Rasch Model (M-RM), Multidimensional 2-Parameter Logistic Model (MD-2PLM), Generalized Linear Model (GLM). \underline{Areas of Research}: \colorbox{green!10}{Psychometrics}, \colorbox{teal!10}{Network Psychometrics}, \colorbox{cyan!10}{Statistics \& ML}, \colorbox{blue!10}{Physics}}
\label{fig:ModelDiagram}
\end{figure*}

\subsection{From binary to continuous variables}
The IM is in more general statistical terms a \textit{Probabilistic Graphical Model} (PGM; \cite{Koller.2009.PGM, Murphy.2012, Chen.2024, Maasch.2025.PGM}), or in that context more specifically a \textit{Pairwise Markov Random Field} (PMRF; \cite{HernandezLemus.2021.MRF}) with binary random variables. From this perspective it has been shown in \textcite[p. 668]{Murphy.2012} that it can be related to the 
GGM\footnote{also referred to as "covariance selection models", "concentration graphs" or "conditional independence graphs" in \textcite{Zerenner.2014.GGM}} a PGM with gaussian random variables or Gaussian Markov Random Field (GMRF; \cite{Rue.2005.GMRF}).

\paragraph{GGM}
Therefore, let us assume now that our dataset consists of observations from $N$ individuals at one time point (T=1) for $P$ centered and normally distributed variables: $\bm{X} \sim \mathcal{N}_P(\bm{0}, \bm{\Sigma})$. The GGM models the pairwise interaction between variables, which means all the relevant information of how the variables are related is contained in $\bm{\hat{\Sigma}}$, usually its inverse, the precision\footnote{\textit{S1.1.2 Precision (prec)} in the library of 237 statistics of pairwise interactions presented in \parencite{Cliff.2023.SPI}} (or concentration) matrix $\bm{K}$ is estimated from the data. A zero entry in $\bm{K}$ is equivalent to stating that two variables are independent given all the other variables (\ref{eq_3.1}).
\begin{align}
\bm{K} &= \bm{\hat{\Sigma}^{-1}} \tag{3} \label{eq_3}\\
(x_i \Perp x_j | \bm{x}^{-(i,j)}) &\Leftrightarrow k_{i,j} = k_{j,i} = 0 \tag{3.1} \label{eq_3.1}\\
cor(X_i, X_j | \bm{x}^{-(i,j)}) &= \frac{\kappa_{ij}}{\sqrt{\kappa_{ii} \kappa_{jj}}} = \omega_{i,j} = \omega_{j,i} \tag{3.2} \label{eq_3.2}
\end{align}
Now $\bm{K}$ encodes the conditional-independence between variables, and can be standardized to contain partial-correlation coefficients (\ref{eq_3.2}). Where $\kappa_{ij}$ are elements in $\bm{K}$, and $\bm{x}^{-(i,j)}$ is the set of variables without i and j. The partial-correlation coefficients are entries in the $P \times P$ weight matrix $\bm{\Omega}$ (sometimes referred to as "Partial-Correlation Network") and can be graphically displayed as a weighted network, with each variable $X_i$ represented as a node and connections (edges) between nodes parametrized by partial-correlations (see Figure \ref{fig:GGM}), compared to IM in Figure \ref{fig:IsingLatticeGraph} it includes additional edges because variables are no longer modeled on a square lattice graph.
\begin{figure}[!htbp]
    \centering
    \begin{tikzpicture}[>=stealth, node distance=2cm]
    \node[rectangle, draw, minimum size=1.2cm] (A) {$X_1$};
    \node[rectangle, draw, below=of A, minimum size=1.2cm] (B) {$X_2$};
    \node[rectangle, draw, right=of B, minimum size=1.2cm] (C) {$X_3$};
    \node[rectangle, draw, right=of A, minimum size=1.2cm] (D) {$X_4$};
    \draw[-] (A) -- node[midway, left] {$\omega_{12}$} (B);
    \draw[-] (B) -- node[midway, below] {$\omega_{23}$} (C);
    \draw[-] (A) -- node[midway, above] {$\omega_{14}$} (D);
    \draw[-] (D) -- node[midway, right] {$\omega_{34}$} (C);
    \draw[-] (A) -- node[near start, left] {$\omega_{13}$} (C);
    \draw[-] (B) -- node[near start, right] {$\omega_{24}$} (D);
    \fill[black] ($(A)+(0,0.8)$) circle (1pt);
    \fill[black] ($(A)+(0,1)$) circle (1pt);
    \fill[black] ($(A)+(0,1.2)$) circle (1pt);

    \fill[black] ($(D)+(0,0.8)$) circle (1pt);
    \fill[black] ($(D)+(0,1)$) circle (1pt);
    \fill[black] ($(D)+(0,1.2)$) circle (1pt);

    \fill[black] ($(D)+(0.8,0)$) circle (1pt);
    \fill[black] ($(D)+(1,0)$) circle (1pt);
    \fill[black] ($(D)+(1.2,0)$) circle (1pt);

    \fill[black] ($(C)+(0.8,0)$) circle (1pt);
    \fill[black] ($(C)+(1,0)$) circle (1pt);
    \fill[black] ($(C)+(1.2,0)$) circle (1pt);
    \end{tikzpicture}
    \caption{The GGM visualized, with $X_p$ being continuous (gaussian) random variables, and $\omega_{i j}$ entries in the in the $P \times P$ weight matrix $\bm{\Omega}$.}
    \label{fig:GGM}
\end{figure}

Since the GGM therefore simply involves standardizing the precision matrix, the following reparameterization (\ref{eq_3a}) has been proposed in \parencite{Epskamp.2017.LNM-RNM, vanBork.2021}.
\begin{equation}
\hat{\bm{\Sigma}} = \hat{\bm{K}}^{-1} = \bm{\Delta} (\bm{I} - \bm{\Omega})^{-1} \bm{\Delta} \tag{3a} \label{eq_3a}
\end{equation}
With $\bm{\Delta}$ a diagonal matrix of $\delta_{ii} = \kappa_{ii}^{-\frac{1}{2}}$, and $\bm{\Omega}$ with zeros on the diagonal and $\omega_{i,j}$ as off-diagonal elements encoding the network structure (edges between variables). This model allows for confirmatory testing and model comparisons with GGMs \parencite{Epskamp.2017.LNM-RNM}.
\paragraph{GGM --- Latent Variables}
As in the binary case, there is a close relation between psychometric latent variable models and PGMs in the continuous (gaussian) case as well. If we estimate a GGM on data that is generated by a univariate latent variable model, there should be no zero elements in $\Omega$, representing a saturated (or "fully-connected") GGM \parencite{vanBork.2021, Waldorp.2022}. Or more generally, data assumed to covary due to some latent variable(s), implies cluster(s) in probabilistic graphical models \parencite{Golino.2017.EGA, Golino.2020.EGA}.

\begin{figure*}[!htbp]
\centering
\begin{tikzpicture}[
    font=\small,
    model/.style={
        rectangle,
        thick,
        minimum width = 0.1cm,
        minimum height = 0.1cm,
        align = center,
        inner sep=1pt
    },
    binary-categorical/.style={
        draw,
        rectangle,
        dashed, ultra thick,
        minimum width = 0.1cm,
        minimum height = 0.1cm,
        align = center,
        inner sep=1pt
    },    
    physics_cross/.style={model, fill=blue!10},
    physics_time/.style={model, fill=blue!45},
    ML_cross/.style={model, fill=cyan!10},
    ML_time/.style={model, fill=cyan!45},
    Graph_cross/.style={model, fill=magenta!15},
    Graph_time/.style={model, fill=magenta!45},
    psychometrics_cross/.style={model, fill=green!10},
    psychometrics_time/.style={model, fill=green!45},
    network_cross/.style={model, fill=teal!10},
    network_time/.style={model, fill=teal!45},
    econometrics/.style={model, fill=orange!45},
    relation/.style={-{Latex[length=2mm]}, thick}
]
\node[ML_cross] (MRF) at (-7.2,5.8) {
\textbf{Markov Random Field (MRF)}
};

\node[ML_cross] (BN) at (-3,6.8) {
\textbf{Bayesian Network (BN)}
};

\node[ML_cross] (GGM) at (-7.2,4) {
\textbf{Gaussian Graphical Model (GGM)}\\[1pt]
${\bm{\Sigma}} = {\bm{K}}^{-1} = \bm{\Delta} (\bm{I} - \bm{\Omega})^{-1} \bm{\Delta}$
};

\node[ML_cross] (GLM) at (-1,2.3) {
\textbf{Generalized Linear Model (GLM)}
};

\node[psychometrics_cross] (NLFM) at (-2,1.2) {
\textbf{NLFM}\\[1pt]
$\Rightarrow \bm{\Sigma} = \bm{\Lambda} \bm{\Lambda}^{\top} + \bm{\Psi}$
};

\node[psychometrics_cross] (PA) at (0,6) {
\textbf{Path Analysis}
};

\node[psychometrics_cross] (SEM) at (0,4) {
\textbf{Structural Equation Model (SEM)}\\[1pt]
$\Rightarrow {\bm{\Sigma}}=\bm{\Lambda}(\bm{I}-\bm{B})^{-1} \bm{\Psi}(\bm{I}-\bm{B})^{-1 \top} \bm{\Lambda}^{\top}+\bm{\Theta}$
};

\node[network_cross] (LNM/RNM) at (-3.7,3) {
\textbf{LNM/RNM}
};

\node[network_cross] (EGM) at (-9,2) {
\textbf{EGM}
};

\node[network_cross] (EGA) at (-7.2,2) {
\textbf{EGA}
};
\node[network_time] (DynEGA) at (-7.2,1) {
\textbf{DynEGA}
};

\draw[relation] (SEM) -- (PA) node[fill=white, midway] {$\bm{\Lambda}=\bm{I}; \bm{\Theta}=\bm{0}$};
\draw[relation] (SEM) -- (GLM);
\draw[relation] (GLM) -- (NLFM);
\draw[] (GGM) -- (NLFM.west) node[fill=white, midway, sloped, font=\small] {Woodbury identity};
\draw[relation] (GGM) -- (EGA) node[fill=white, midway, font=\small] {+ walktrap};
\draw[relation] (EGM) -- (EGA);
\draw[relation] (EGA) -- (DynEGA);
\draw[relation] (GGM) -- (LNM/RNM.west);
\draw[relation] (SEM) -- (LNM/RNM.east);
\draw[relation] (MRF) -- (GGM) node[fill=white, midway, font=\small] {undirected \& Gaussian};
\draw[relation] (MRF.east) -- (BN.south) node[fill=white, midway, font=\small] {DAG};

\end{tikzpicture}
\caption{Latent Network Modeling (LNM), Residual Network Modeling (RSM), Exploratory Graph Analysis (EGA), Dynamic EGA (DynEGA), Exploratory Graph Model (EGM). \underline{Areas of Research}: \colorbox{green!10}{Psychometrics}, Network Psychometrics (\colorbox{teal!10}{cross-sectional} / \colorbox{teal!45}{time-series}), \colorbox{cyan!10}{Statistics \& ML}, \colorbox{blue!10}{Physics}}
\label{fig:ModelDiagram_2}
\end{figure*}

\subsection{Latent Variable Models: CFA, SEM, GGM, and Bayesian Networks (Figure \ref{fig:ModelDiagram_2})}
Therefore, let's have a closer look at latent variable models for continuous (gaussian) variables. If the correlations in $\bm{X}$ are assumed to be explained by $M$ (centered) continuous\footnote{Connections between the binary FA model and the two-parameter IRT model have been established as well \parencite[see][]{Takane.1987.FM-IRT, Kamata.2008.FM-IRT}.} latent variables $\bm{\eta}$ (local independence: $X_i \Perp X_j \mid \eta$), we are in the world of \textit{Confirmatory Factor Analysis} (CFA; \cite{Wang.2021}): 
\begin{align}
    \bm{x} &= \bm{\Lambda} \bm{\eta} + \bm{\varepsilon} \tag{4} \label{eq_4}\\
    &\Rightarrow \bm{\hat{\Sigma}}_{CFA} = \bm{\Lambda} \bm{\Psi}^\eta \bm{\Lambda}^\top + \bm{\Theta}^\varepsilon \tag{4.1} \label{eq_4.1}
\end{align}
With $\bm{\eta} \sim \mathcal{N}_M(\bm{0}, \bm{\Psi})$, and $\bm{\varepsilon} \sim \mathcal{N}_P(\bm{0}, \bm{\Theta})$ (conditional independence: $\eta \Perp \varepsilon$ $\Rightarrow$ diagonal $\bm{\Theta})$.

\paragraph{Factor Analysis --- GGM}
Equation (\ref{eq_4.1}) can be inverted to obtain an expression for the equivalent GGM, the precision matrix $\bm{K}$ then becomes a block diagonal matrix in which every block is constructed of the inner product of factor loadings and inverse residual variances \parencite[see][for details]{Golino.2017.EGA}.

\begin{figure}[!htbp]
    \centering
    \begin{tikzpicture}[>=stealth, node distance=2cm]
    \node[circle, draw, minimum size=1.2cm] (eta1) {$\eta_1$};
    \node[circle, draw, right=2cm of eta1, minimum size=1.2cm] (eta2) {$\eta_2$};
    \node[rectangle, draw, below left=1.5cm and 0.1cm of eta1, minimum width=1cm, minimum height=1cm] (x1) {$X_1$};
    \node[rectangle, draw, below right=1.5cm and 0.1cm of eta1, minimum width=1cm, minimum height=1cm] (x2) {$X_2$};
    \node[rectangle, draw, below left=1.5cm and 0.1cm of eta2, minimum width=1cm, minimum height=1cm] (x3) {$X_3$};
    \node[rectangle, draw, below right=1.5cm and 0.1cm of eta2, minimum width=1cm, minimum height=1cm] (x4) {$X_4$};
    \draw[->] (eta1) -- node[left] {$\lambda_{11}$} (x1);
    \draw[->] (eta1) -- node[right] {$\lambda_{21}$} (x2);
    \draw[->] (eta2) -- node[left] {$\lambda_{32}$} (x3);
    \draw[->] (eta2) -- node[right] {$\lambda_{42}$} (x4);
    \draw[<->, bend left=20] (eta1) to node[above] {$\psi_{21}^{\eta}$} (eta2);
    \node[above left=0.1cm and -0.2cm of eta1] {$\psi_{11}^{\eta}$};
    \node[above=0.1cm of eta2] {$\psi_{22}^{\eta}$};
    \node[below=0.4cm of x1] (d1) {$\varepsilon_1$};
    \node[below=0.4cm of x2] (d2) {$\varepsilon_2$};
    \node[below=0.4cm of x3] (d3) {$\varepsilon_3$};
    \node[below=0.4cm of x4] (d4) {$\varepsilon_4$};
    \foreach \i in {1,2,3,4}{
    \draw[->] (d\i) -- (x\i);
    }
    \node[left=0.1cm of d1] {$\theta^{\varepsilon}_{11}$};
    \node[left=0.1cm of d2] {$\theta^{\varepsilon}_{22}$};
    \node[right=0.1cm of d3] {$\theta^{\varepsilon}_{33}$};
    \node[right=0.1cm of d4] {$\theta^{\varepsilon}_{44}$};
    \end{tikzpicture}
    \caption{Illustration of a CFA model with four observed variables ($X_1, X_2, X_3, X_4$) and two latent variables ($\eta_1, \eta_2$).}
    \label{fig:CFA}
\end{figure}

\paragraph{Structural Equation Modeling}
In SEM relations between observations can additionally be considered, which extends Equation (\ref{eq_4}) by a square matrix $\bm{B}$ with zeros on the diagonal (\ref{eq_4a}), and (potentially) causal effects between observations on the off-diagonal \parencite{Epskamp.2018.GGM-GVAR}.
\begin{equation}
    \bm{x} = \bm{B} \bm{x} + \bm{\Lambda} \bm{\eta} + \bm{\varepsilon} \tag{4a} \label{eq_4a}
\end{equation}
Furthermore, additional structure\footnote{SEM has also been related to \textit{Multilevel Models} \parencite[see][]{Curran.2003.SEM-MLM}.} can be imposed on $\bm{\eta}$ by including structural linear relations between latent variables (\ref{eq_4a.1}). And for $\bm{\Lambda} = \bm{0}, \bm{\Theta} = \bm{0}$ the model reduces to Path Analysis \parencite{Wright.1934, Epskamp.2017.LNM-RNM}. 
\begin{equation}
    \bm{\eta} = \bm{B} \bm{\eta} + \bm{\zeta}   \tag{4a.1} \label{eq_4a.1}
\end{equation}
\begin{equation}
    \Rightarrow \bm{\hat{\Sigma}}_{SEM} = \bm{\Lambda} (\bm{I} - \bm{B})^{-1} \bm{\Psi}^\eta (\bm{I} - \bm{B})^{-1 \top} \bm{\Lambda}^{\top} + \bm{\Theta}^\zeta \tag{4a.1.1} \label{eq_4a.1.1}
\end{equation}
For Equation (\ref{eq_4.1}) or (\ref{eq_4a.1.1}) to be identifiable in practice, usually many restrictions need to be imposed, for example, $\bm{\Theta}$ to be diagonal (local independence assumption) but also on $\bm{\Lambda}$. To improve model fit some off-diagonal elements of $\bm{\Theta}$ can be estimated, but this is an uncontrolled and ad-hoc process and systematic violations of local independence cannot be accounted for \parencite{Epskamp.2017.LNM-RNM}. Many elements in $\bm{\Theta}$ should ideally be freely estimated.
\paragraph{LNM \& RNM}
In \cite{Epskamp.2017.LNM-RNM} the SEM framework has been extended by incorporating a GGM to either model the conditional independence between latent variables (Equation (\ref{eq_4.1}) but with Equation (\ref{eq_3a}) for $\bm{\Psi}$) which is termed \textit{Latent Network Modeling} (LNM) and represents a good method for initial exploratory estimation. Or \textit{Residual Network Modeling} (RNM) which models "residual interactions" (Equation (\ref{eq_4a.1.1}) but with Equation (\ref{eq_3a}) for $\bm{\Theta}$), which allows for the estimation of a factor structure, while being able to model violations of local independence, but still have some structure on the error-correlations in the form of the "residual network".

We can roughly summarize by saying that "network models" isolate variance that is unique to variable pairs, while latent variable models focuses on variance that is shared across all variables. But they are closely related in that they both model a constrained covariance structure. Therefore, latent variable modeling and "network modeling" can be viewed from a purely statistical perspective much more productively as complementing each other instead of excluding or replacing the other \parencite{Bringmann.2018.Psychopathology}. For example, SEM usually requires us to impose more structure than we would perhaps want to for exploratory estimation, and the assumption of no latent variables in GGMs might not be so desirable as well. Initial exploratory factor analysis could therefore be done by checking for clusters in a GGM to inform the number of factors \parencite{Golino.2017.EGA, Golino.2020.EGA}. But it is important to keep in mind that the conventionally used estimation methods \parencite{Epskamp.2017.Estimation, Isvoranu.2023.Estimation} try to estimate a sparse GGM (i.e., some elements in $\bm{K}$ constrained to be exact zeroes), which is only expected given a factor model when latent variables are orthogonal \parencite{Epskamp.2018.GGM-GVAR}. Moreover, in \cite{Christensen.2021, Christensen.2025} a network equivalent of factor loadings, which they call "network loadings" is proposed, and the applicability of SEM fit indices in \textit{Confirmatory Network Analysis} (CNA; \cite{Du.2025.panelGVAR/CNA/CFA}) has been evaluated.

\paragraph{GGM --- SEM --- DAG (Causal Inference)}
When we only consider the causal model (Equation (\ref{eq_4a}) without latent variables) the following decomposition of the covariance-matrix is implied:
\begin{equation}
    \Rightarrow \bm{\hat{\bm{\Sigma}}} = (\bm{I} - \bm{B})^{-1} \bm{\Theta}^\varepsilon (\bm{I} - \bm{B})^{-1 \top} \tag{4b.1} \label{eq_4b.1}
\end{equation}
Which can be inverted as well, to represent a GGM in which edges are indicative of potential causal effects \parencite{Epskamp.2018.GGM-GVAR}. In causal-inference research similarities as well as differences between the GGM, SEM and Directed Acyclic\footnote{no cycles or feedback loops between variables} Graph (DAG) have been noted as well \parencite{Ryan.2022.SEM-DAG-GGM, Kunicki.2023.SEM-DAG}. Most generally again in statistical terms, a DAG (also known as \textit{Bayesian Network} (BN; \cite{Murphy.2012})) is a PGM with unweighted but directed edges between variables \parencite{Ryan.2022.SEM-DAG-GGM}. Therefore, like the GGM and IM are instances of an undirected PMRF, a Bayesian Network is an instance of a directed PMRF. And when the causal relations between variables are assumed to be linear with gaussian residuals, the causal system can be represented as a SEM \parencite{Ryan.2022.SEM-DAG-GGM}. But, causal structure is generally not uniquely identifiable from only observational data, even if the causal relations are assumed to be a linear SEM (issue of factor score indeterminacy\footnote{„the fatal flaw that factor indeterminacy reveals for the common factor model, is fatal only for the unreasonable expectations we have for this model, especially in its exploratory applications [...] What is unreasonable is to expect common factor analysis to produce from data unambiguous, self-evident insights into the workings of the world. In fact, I believe we must abandon the belief that there are any such methods in science. The belief that there are such methods was a delusion of the empiricists whose views had a very strong influence on the development of exploratory statistics.“ \parencite{Mulaik.1986.FactorIndeterminacy}} \parencite[see][]{Waller.2023.FactorIndeterminacy}).

\subsection{Time-Series Analysis: Psychometrics meets Econometrics (Figure \ref{fig:ModelDiagram_3})}
\begin{figure*}[!htbp]
\centering
\begin{tikzpicture}[
    font=\small,
    model/.style={
        rectangle,
        thick,
        minimum width = 0.1cm,
        minimum height = 0.1cm,
        align = center,
        inner sep=1pt
    },
    binary-categorical/.style={
        draw,
        rectangle,
        dashed, ultra thick,
        minimum width = 0.1cm,
        minimum height = 0.1cm,
        align = center,
        inner sep=1pt
    },    
    physics_cross/.style={model, fill=blue!10},
    physics_time/.style={model, fill=blue!45},
    ML_cross/.style={model, fill=cyan!10},
    ML_time/.style={model, fill=cyan!45},
    Graph_cross/.style={model, fill=magenta!15},
    Graph_time/.style={model, fill=magenta!45},
    psychometrics_cross/.style={model, fill=green!10},
    psychometrics_time/.style={model, fill=green!45},
    network_cross/.style={model, fill=teal!10},
    network_time/.style={model, fill=teal!45},
    econometrics/.style={model, fill=orange!45},
    relation/.style={-{Latex[length=2mm]}, thick}
]
\node[ML_cross] (GGM) at (-12,0) {
\textbf{GGM}
};

\node[psychometrics_cross] (SEM) at (-1,0) {
\textbf{SEM}
};

\node[network_time] (gVAR) at (-12,-2) {
\textbf{gVAR}
};

\node[network_time] (multi-VAR) at (-10,0) {
\textbf{multi-VAR}
};

\node[network_time] (ALSVAR) at (-10,-1) {
\textbf{ALS VAR}
};

\node[network_time] (fitlandr) at (-7,-1) {
\textbf{fitlandr}
};

\node[econometrics] (VAR) at (-7,-2) {
\textbf{VAR(1)}\\[1pt]
$\bm{y}_t = \bm{B} \bm{y}_{t-1}+ \bm{\varepsilon}_t$
};

\node[psychometrics_time] (uSEM) at (-4.5,0) {
\textbf{uSEM}
};

\node[network_time] (GIMME) at (-7,0) {
\textbf{GIMME}
};

\node[econometrics] (SVAR) at (-2,-1) {
\textbf{Structural VAR}
};

\node[econometrics] (VDAR) at (-2,-2) {
\textbf{VDAR}
};

\node[network_time] (lvgvar) at (-12,-3) {
\textbf{ts-lvgvar}
};

\node[psychometrics_time] (NDLC-SEM) at (-12,-4) {
\textbf{NDLC-SEM}
};

\node[network_time] (mlVAR) at (-10,-3) {
\textbf{mlVAR}
};

\node[psychometrics_time] (DSEM) at (-7,-4) {
\textbf{DSEM}
};

\node[psychometrics_time] (CLPM) at (-7,-3) {
\textbf{CLPM}
};

\node[psychometrics_time] (DFM) at (-3,-4) {
\textbf{DFM}
};

\node[psychometrics_time] (P) at (-3,-3) {
\textbf{P-Technique}
};

\node[psychometrics_time] (DLCA) at (-12,-5.5) {
\textbf{DLCA}
};

\node[ML_time] (HMM) at (-5,-5.5) {
\textbf{Hidden Markov Model (HMM)}
};

\draw[relation] (gVAR) -- (lvgvar);
\draw[relation] (gVAR) -- (GGM) node[fill=white, midway] {$\bm{B} = \bm{0}$};
\draw[relation] (VAR) -- (gVAR);
\draw[relation] (VAR.west) -- (mlVAR.east);
\draw[relation] (VAR) -- (multi-VAR.south);
\draw[relation] (VAR.west) -- (ALSVAR);
\draw[dashed] (VAR) -- (fitlandr);
\draw[relation] (VAR) -- (SVAR.west) node[fill=white, midway] {$+ \bm{A} \bm{y}_t $};
\draw[relation] (VAR) -- (VDAR) node[fill=white, midway, sloped, font=\small] {Binary};
\draw[dashed] (VAR) -- (CLPM);
\draw[relation] (DFM) -- (HMM);
\draw[relation] (DFM) -- (DSEM) node[fill=white, midway,  font=\small] {multilevel};
\draw[relation] (DFM) -- (P);
\draw[relation] (SVAR) -- (uSEM.south);
\draw[relation] (uSEM) -- (GIMME);
\draw[relation] (SEM) -- (uSEM) node[fill=white, midway] {$T>1$};
\draw[dashed] (lvgvar) -- (NDLC-SEM);
\draw[relation] (DSEM) -- (NDLC-SEM) node[fill=white, midway] {$\Omega_2 h_2\left(\eta_{2 i}\right)$};
\draw[dashed] (DSEM) -- (CLPM);
\draw[relation] (DSEM.west) -- (mlVAR.east);
\draw[relation] (DSEM.west) -- (DLCA.east);
\draw[relation] (HMM) -- (DLCA.east) node[fill=white,  pos=0.4] {$\left[Y_{1, i t} \mid S_{i t}\right]$};
\draw[relation] (DLCA) -- (NDLC-SEM) node[fill=white, midway] {$\alpha_{itdc}$};

\end{tikzpicture}
\caption{Gaussian Graphical Model (GGM), Vector Autoregressive (VAR), graphical VAR (gVAR), time-series latent variable gVAR (ts-lvgvar), multilevel VAR (mlVAR), Vector Discrete AR (VDAR), Structural Equation Modeling (SEM), unified SEM (uSEM), Dynamic Factor Model (DFM), Cross-Lagged Panel Model (CLPM), Dynamic Structural Equation Modeling (DSEM), Dynamic Latent Class Analysis (DLCA), Nonlinear Dynamic Latent Class - Structural Equation Model (NDLC-SEM).  \underline{Areas of Research}: Psychometrics ( \colorbox{green!10}{cross-sectional} / \colorbox{green!45}{time-series}), \colorbox{teal!45}{Network Psychometrics}, \colorbox{orange!45}{Econometrics}, Statistics \& ML (\colorbox{cyan!10}{cross-sectioanl} / \colorbox{cyan!45}{time-series})}
\label{fig:ModelDiagram_3}
\end{figure*}

Moving from the cross-sectional into the time-series domain is usually done by incorporating models and ideas from econometrics, where one of the most basic approaches for analyzing multivariate time-series data is the \textit{Vector-Autoregressive} (VAR; \cite{Lutkepohl.2003.VAR, Stock.2001.VAR, Duker.2025.VARMA}) model in Equation (\ref{eq_5}). For notational simplicity, in the following we will only consider lag-1 (autoregressive) relationships between variables, but higher order lags are common as well and the ARMA model family in general is much bigger \parencite{Holan.2010.ARMA}.
\begin{equation}
    \bm{x}_t = \bm{B} \bm{x}_{t-1} + \bm{\varepsilon}_t \quad \text{with} \quad \bm{\varepsilon}_T \sim \  \mathcal{N}(\bm{0}, \bm{\Theta})   \tag{5} \label{eq_5}
\end{equation}
Where we assume that our data, response of subject $n$ on all $P$ items at time point $t$, is $\bm{x}_{t}\sim \mathcal{N}_P(\bm{0}, \bm{\Sigma})$ (with $N$=1, $P$=Large, $T$=Large). The VAR can be viewed as a generalization of the GGM into the time-series domain, where some form of dependence between repeatedly observed variables is expected \parencite{Epskamp.2018.GGM-GVAR}.

\paragraph{GGM + VAR = gVAR}
A GGM can also be used to model the residual (innovation) structure of a VAR model, this is termed \textit{graphical VAR} (gVAR; \cite{Epskamp.2018.GGM-GVAR, Wild.2010.gVAR}) and can be written as a conditional gaussian distribution in Equation (\ref{eq_6}).  
\begin{equation}
\bm{x}_T \mid \bm{x}_{T-1}=\bm{x}_{t-1} \sim \mathcal{N}\left(\bm{B} \bm{x}_{t-1}, \bm{\Theta}\right) \tag{6} \label{eq_6}
\end{equation}
Here $\bm{B}$ is usually referred to as "temporal network" and $\bm{K}^{\bm{\Theta}} = \bm{\Theta}^{-1}$ as the "contemporaneous network" in network psychometrics. With $\boldsymbol{B}=\boldsymbol{0}$, the gVAR model reduces to the GGM in Equation (\ref{eq_3}). Also note that Equation (\ref{eq_5}) and (\ref{eq_6}) are equivalent representations.

\paragraph{VAR --- SVAR --- uSEM}
In the VAR model of Equation (\ref{eq_5}), contemporaneous relations between variables are only considered in $\varepsilon_t$, which is in the gVAR modeled more explicitly as a GGM. It is therefore interesting to highlight that the VAR is a "reduced form" of a more general class of models known as the \textit{Structural VAR} (SVAR; \cite{Chen.2011.SVAR}) in Equation (\ref{eq_7}), which combines a SEM for contemporaneous and a VAR for lagged dependencies between variables and therefore explicitly models the directional relations in the "contemporaneous network".
\begin{equation}
    \bm{x}_t = \textbf{A} \bm{x}_t + \bm{B} \bm{x}_{t-1} + \bm{\varepsilon}_t \tag{7} \label{eq_7}
\end{equation}
The sparsity of $\bm{K}^{\bm{\Theta}}$ in a gVAR model corresponds in the same way to the sparsity of the directed "contemporaneous network" in a SVAR model, as the GGM corresponds to DAG's (edges arise in the GGM due to edges in the causal network or conditioning on common effects). However, $\bm{B}$ is sparser in Equation (\ref{eq_7}) than in the gVAR, as contemporaneous mediators can be controlled for in SVAR but not in the gVAR \parencite{Epskamp.2018.GGM-GVAR}. But the common method of estimating SVAR models leads to multiple solutions \parencite{Beltz.2016.GIMME-MS}, to avoid this issue of non-uniqueness, step-wise model selection can be used, for example in the \textit{unified SEM} (uSEM\footnote{Additionally: \textit{extended uSEM} (euSEM; \cite{Gates.2011.euSEM}), \textit{hybrid uSEM} (huSEM; \cite{Ye.2021.uSEM/hybridVAR/huSEM}), \textit{regularized huSEM} (reg-huSEM; \cite{Ye.2021.uSEM/hybridVAR/huSEM})}; \cite{Kim.2007, Gates.2010.uSEM}) framework or with Bayesian Dynamical SEM models \parencite{Asparouhov.2018.DSEM}.

\paragraph{uSEM --- GIMME}
When used for analyzing time-series data of many individuals (ILD $\Rightarrow$ $N$=Large, $T$=Large, $P$=Large), uSEM models are either applied separately to individuals or to aggregated data across individuals. The \textit{Group Iterative Multiple Model Estimation} (GIMME\footnote{Additionally: \textit{Confirmatory Subgroup} (CS-GIMME; \cite{Henry.2019.CS-GIMME}), \textit{Latent Variable} (LV-GIMME; \cite{Gates.2020.LV-GIMME, Ye.2024}), \textit{Continuous-Time} (CT-GIMME; \cite{Park.2024.CT-GIMME}), \textit{hybrid-GIMME} \parencite{Luo.2023.hybrid-GIMME} , and \textit{GIMMEgVAR} \parencite{Lee.2024.GIMMEgVAR}. See also \url{https://tarheels.live/gimme/}}; \cite{Gates.2012.GIMME, Beltz.2016.GIMME-MS, Lane.2017.GIMME, Lee.2026.reviewGIMME}) framework has been proposed and explicitly designed to analyze such data, and can account even for possibly heterogeneous individuals in the dataset. The GIMME framework was initially developed to analyze fMRI data, but is increasingly used for analyzing behavioral and psychological data \parencite{Robinaugh.2020.Psychopathology, Blanchard.2023.Review}. As a likely characteristic of ILD it can be assumed that individuals will present some common but also some unique variation. GIMME therefore starts by identifying relations between variables that hold for the majority of individuals in the dataset, this model is then used as a starting point for estimating individual patterns. The still prevailing issue of potential multiple solutions can be mitigated by \textit{GIMME for Multiple Solutions} (GIMME-MS; \cite{Beltz.2016.GIMME-MS}) in Equation (\ref{eq_7a}).
\begin{align}
    \bm{x}_{i,t} &=  (\bm{A}_i + \bm{A}^{(g)}) \bm{x}_{i,t} \notag\\
    & + (\bm{B}_{i,l} + \bm{B}^{(g)}) \bm{x}_{i,t-1} + \bm{\varepsilon}_{i,t} \tag{7a} \label{eq_7a}
\end{align}
Where $\bm{A}$ is a matrix of regression coefficients with the subscript $i$ indicating individual-level effects and the superscript $g$ for group-level effects. This can be further extended by allowing for subgrouping effects (S-GIMME; \cite{Gates.2017.S-GIMME}) with $k$ as the sub-group index in Equation (\ref{eq_7b}).
\begin{align}
    \bm{x}_{i,t} & = (\bm{A}_i + \bm{A}_{i,k}^{(s)} + \bm{A}^{(g)}) \bm{x}_{i,t} \notag\\
    & + (\bm{B}_{i,l} + \bm{B}_{i,k}^{(s)} + \bm{B}^{(g)}) \bm{x}_{i,t-1} + \bm{\varepsilon}_{i,t} \tag{7b} \label{eq_7b}
\end{align}

\paragraph{SEM + VAR = DSEM}
Similar to what we have just outlined in the network psychometrics context, we can generalize a cross-sectional latent variable model like the SEM into the time-series domain by incorporating lagged effects as well, this is called \textit{Dynamic SEM} (DSEM\footnote{multilevel extension of the \textit{Dynamic Factor Model} (DFM; \cite{Molenaar.1985.DFM, Molenaar.2017.DFM}) }; \cite{Asparouhov.2018.DSEM}). The most general formulation of the DSEM consists of three sets of structural equation models (with the usual measurement and structural equations), on level-1 with $\bm{y}_{i,t}^{(1)}$ as within-person effects, and on level-2 with $\bm{y}_{i}^{(2)}$ as between-person and $\bm{y}_{t}^{(3)}$ for time-specific effects. For simplicity, in the following we will only consider the first two components, which is termed "two-level DSEM"\footnote{Again, we will only considers lag-1 ($l=1$) relationships which simplifies the equations, but modeling higher order lags is possible.}. 
\begin{equation}
    \bm{x}_{i,t} = \bm{y}_{i,t}^{(1)} + \bm{y}_{i}^{(2)} \tag{8} \label{eq_8}
\end{equation}
With $\bm{x}_{i,t}$ a vector of observations for individual $i$ at time $t$, which could be for example affective states measured repeatedly with $P$ items. The within-person effects then have a time series component (\ref{eq_8.1}).
\begin{align}
    \bm{y}_{i,t}^{(1)} &= \bm{\nu}^{(1)} + \bm{\Lambda}_{l}^{(1)} \bm{\eta}_{i,t-l}^{(1)} +  \bm{R}_{l}^{(1)} \bm{y}_{i,t-l}^{(1)} \notag\\
    & + \bm{K}_{l}^{(1)} \bm{v}_{i,t-l}^{(1)} + \bm{\epsilon}_{i,t}^{(1)} \tag{8.1} \label{eq_8.1}\\
    \bm{\eta}_{i,t-l}^{(1)} &= \bm{\alpha}^{(1)} + \bm{B}_{l}^{(1)} \bm{\eta}_{i,t-l}^{(1)} + \bm{Q}_{l}^{(1)} \bm{y}_{i,t-l}^{(1)} \notag\\
    & + \bm\Gamma_{l}^{(1)} \bm{v}_{i,t-l}^{(1)} + \bm\xi_{i,t}^{(1)} \tag{8.1.1} \label{eq_8.1.1}\\
    \bm{y}_{i}^{(2)} &= \bm{\nu}^{(2)} + \bm{\Lambda}^{(2)} \bm{\eta}_{i}^{(2)} + \bm{K}^{(2)} \bm{v}_{i}^{(2)} + \bm{\epsilon}_{i}^{(2)} \tag{8.2} \label{eq_8.2}\\
    \bm{\eta}_{i}^{(2)} &= \bm{\alpha}^{(2)} + \bm{B}^{(2)} \bm{\eta}_{i}^{(2)} + \bm{\Gamma}^{(2)} \bm{v}_{i}^{(2)} + \bm{\xi}_{i}^{(2)} \tag{8.2.1} \label{eq_8.2.1}
\end{align}
With $\bm{v}_{i,t-1}^{(1)}$ and $\bm{\eta}_{i,t-l}^{(1)}$ being vectors of observed covariates and latent variables for individual $i$ at time $t$, and individual-specific but time-invariant effects in (\ref{eq_8.2}), $\bm{\epsilon}_{i}^{(2)}, \bm{\xi}_{i}^{(2)}$ are zero mean residuals, the remaining vectors and matrices are non-random model parameters. In practice however many restrictions need to be imposed for identification purposes \parencite[see][for details]{Asparouhov.2018.DSEM}. The DSEM framework can model heterogeneity of the intra-individual trajectories by individual-specific differences and time-specific effects as random effects, but unobserved discrete states might be another source of heterogeneity in the data which cannot be considered in this framework.

\paragraph{DSEM + HMM = DLCA}
This shortcoming is addressed in \textit{Dynamic Latent Class Analysis} (DLCA; \cite{Asparouhov.2017.DLCA}), it adds discrete latent states (also referred to as "latent classes" or "regimes") and models the dynamics of these states as a \textit{Hidden Markov Model} (HMM\footnote{can be viewed as a time-series generalization of BNs \parencite[see][]{Ghahramani.2001.HHM-BN}}; \cite{Helske.2019.HMM}). The parameters in Equation (\ref{eq_8.1}) now depend on a latent state $S_{i,t} = s$ for individual $i$ at time $t$, and let's just write this in Equation (\ref{eq_9.1}) only considering the latent variables and observed covariates for simplicity. The latent class variable $S_{i,t}$ changes with a person-specific transition probability modeled as random effects (\ref{eq_9.2}). Such a latent state could represent the intention to be compliant with some treatment, which is usually a variable that is unobserved (or unobservable) but nonetheless has some effects on the outcome variable (e.g., patterns in depression dynamics).
\begin{align}
    [\bm{y}_{i,t}^{(1)} | &S_{i,t} = s] = \bm{\nu}_{s}^{(1)} + \bm{\Lambda}_{l,s}^{(1)} \bm{\eta}_{i,t-l}^{(1)} + \bm\epsilon_{i,t}^{(1)} \tag{9.1} \label{eq_9.1}\\
    [\bm{\eta}_{i,t-l}^{(1)} | &S_{i,t}=s] = \bm{\alpha}_{s}^{(1)} + \bm{B}_{l,s}^{(1)} \bm{\eta}_{i,t-l}^{(1)} &\notag\\
    &+ \bm{\Gamma}_{l,s}^{(1)} \bm{v}_{i,t-l}^{(1)} + \bm{\xi}_{i,t}^{(1)} \tag{9.1.1} \label{eq_9.1.1}\\
    P(S_{i,t}=d | &S_{i,t-1}=c) = \dfrac{exp(\alpha_{idc})}{\sum_{k=1}^{K} exp(\alpha_{ikc})} \tag{9.2} \label{eq_9.2}
\end{align}
The equations for the between-person effects (\ref{eq_8.2}) stay the same, $\bm{\eta}_{i}^{(2)}$ contains all subject-specific random effects, now including the random transition probability effects $\bm{\alpha}_{idc}$, as well as all random intercepts, loadings, and slopes.

\paragraph{DLCA --- NDLC-SEM}
The DLCA framework has been further extended with \textit{Nonlinear Dynamic Latent Class Structural Equation Modeling} (NDLC-SEM\footnote{See \textcite{Kelava.2022.NDLC-SEM} for an empirical example and \textcite{Faleh.2026.NDLC-SEM_Tutorial} for a detailed, tutorial-like treatment of the dynamic latent variable framework.}; \cite{Kelava.2019.NDLC-SEM, Andriamiarana.2023.NDLC-SEM}) to include flexible non-linear effects between the latent variables (\ref{eq_8.2.1a}), and transition probabilities are additionally dependent on time-specific variables (\ref{eq_9.2a}).
\begin{align}
    \bm{\eta}_{i}^{(2)} = \bm{\alpha}^{(2)} + \bm{B}^{(2)} \bm{\eta}_{i}^{(2)} +\bm{\Omega}^{(2)}h^{(2)}(\bm{\eta}_{i}^{(2)}) \notag\\ 
    + \bm{\Gamma}^{(2)} \bm{v}_{i}^{(2)} + \bm{\xi}_{i}^{(2)} \tag{8.2.1a} \label{eq_8.2.1a}\\
    P(S_{i,t}=d | S_{i,t-1}=c) = \dfrac{exp(\alpha_{itdc})}{\sum_{k=1}^{K} exp(\alpha_{itkc})} \tag{9.2a} \label{eq_9.2a}
\end{align}
Where $h^{(2)}(\bm{\eta}_{i}^{(2)})$ is a vector of functions of $\bm{\eta}_{i}^{(2)}$, and can be used very flexibly with splines to approximate complex relationships between the latent variables. In the within-level model (\ref{eq_8.1}) such non-linear effects can be included as well.

The above mentioned dynamic latent variable models can account for many complex relationships in the data, but due to their factor analytic origin they are highly parameterized models with a lot of structure. And as in the cross-sectional domain, in the time-series domain this might be undesirable as well for initial exploratory estimation.
In \cite{Epskamp.2020.panel/ts-lvgvar} a general framework for modeling relationships between latent variables as a GGM in both time-series data of a single subject (\textit{ts-lvgvar}; $N$=1, $T$=Large) and in panel data (\textit{panel-lvgvar}\footnote{\textit{panelGVAR} in \parencite{Du.2025.panelGVAR/CNA/CFA}}; $T$=3-6) has been proposed. When the contemporaneous relationships are not modeled as a GGM, the ts-lvgvar can be seen as a DFM, and when all variables in the network are observed it reduces to a gVAR model.

\subsection{Novel Directions (Figure \ref{fig:ModelDiagram_4})}
\begin{figure*}[!htbp]
\centering
\begin{tikzpicture}[
    font=\small,
    model/.style={
        rectangle,
        thick,
        minimum width = 0.1cm,
        minimum height = 0.1cm,
        align = center,
        inner sep=1pt
    },
    binary-categorical/.style={
        draw,
        rectangle,
        dashed, ultra thick,
        minimum width = 0.1cm,
        minimum height = 0.1cm,
        align = center,
        inner sep=1pt
    },    
    physics_cross/.style={model, fill=blue!10},
    physics_time/.style={model, fill=blue!45},
    ML_cross/.style={model, fill=cyan!10},
    ML_time/.style={model, fill=cyan!45},
    Graph_cross/.style={model, fill=magenta!15},
    Graph_time/.style={model, fill=magenta!45},
    psychometrics_cross/.style={model, fill=green!10},
    psychometrics_time/.style={model, fill=green!45},
    network_cross/.style={model, fill=teal!10},
    network_time/.style={model, fill=teal!45},
    econometrics/.style={model, fill=orange!45},
    relation/.style={-{Latex[length=2mm]}, thick}
]
\node[physics_cross] (IM) at (-6,1.5) {
\textbf{Ising Model (IM)}\\[1pt]
$p(\bm{x}) \propto \exp\!\left( \sum_i^{n} \mu_i x_i + \sum_{<i,j>} \sigma_{ij} x_i x_j\right)$
};

\node[network_cross] (IdiographicIM) at (-6,4) {
\textbf{Idiographic IM}\\[1pt]
$p(\bm{x}) = \sum_{\bm{w}} \overbrace{p(\bm{x} \mid \bm{w})}^{\text {graph coloring }} p(\bm{w})$
};

\node[network_cross] (DCM) at (-1,4) {
\textbf{Divide \& Color}\\[1pt]
$p(\bm{x}) = \sum_{\bm{w}} \overbrace{p(\bm{x} \mid \bm{w})}^{\text {graph coloring }} p(\bm{w})$
};

\node[Graph_cross] (ER) at (1,6.35) {
\textbf{Erd\H{o}s--R\'enyi}
};

\node[Graph_cross] (RCM) at (-6,6.4) {
\textbf{Random Cluster Model}\\[1pt]
$p(\bm{w}) \propto \prod_{i=1}^{n-1} \prod_{j=i+1}^n \theta_{i j}^{w_{i j}}\left(1-\theta_{i j}\right)^{1-w_{i j}} \lambda^{\kappa(\bm{w})}$
};

\draw[relation] (ER.south) -- (DCM.north) node[fill=white, midway] {$p(\bm{w})$};
\draw[relation] (RCM) -- (IdiographicIM) node[fill=white, midway] {$p(\bm{w})$};
\draw[relation] (RCM) -- (ER) node[fill=white, midway] {$\lambda^{\kappa(\bm{w})}=1$};
\draw[] (IM) -- (IdiographicIM.south) node[fill=white, midway] {$\sigma_{i j} = -\frac{1}{2} \log \left(1-\theta_{i j}\right)$};
\draw[relation] (RCM) -- (ER) node[fill=white, midway] {$\lambda^{\kappa(\bm{w})}=1$};

\end{tikzpicture}
\caption{Erd\H{o}s--R\'enyi (ER or Random Graph Model). \underline{Areas of Research}: \colorbox{teal!10}{Network Psychometrics}, \colorbox{blue!10}{Physics}, \colorbox{magenta!10}{Graph Theory \& Network Science}}
\label{fig:ModelDiagram_4}
\end{figure*}

More recently some very interesting additional relations involving the IM could be established. So, let's go back to the beginning again and considering a fundamental assumption of the IM, in the cross-sectional context we assume that the rows of $\bm{X}$ are $N$ independent identically distributed (i.i.d.) realizations from a IM. For psychological data this seems implausible, it is likely more accurate to assume there is some heterogeneity in the sample (i.e., observations are samples from several different subpopulations exhibiting different symptomatic patterns) and not considering this may lead to the estimation of model parameters that don't represent any of the individuals or perhaps only a very small subset \parencite{Brusco.2019.Ising}. This implies that the IM can only describe a network structure which is invariant across individuals, but in \textcite{Marsman.2023b.IdiographicIsing} it has been shown that this is not necessarily the case.

The \textit{encompassing network model} (or Idiographic IM; \cite{Fortuin.1972, Marsman.2023.IdiographicIsing}) simultaneously models the topology of the network as a random graph\footnote{Random Graph \parencite{Frieze.2015.RandomGraphs} is a general term for probability distributions over graphs, usually it refers to the Erdős-Rényi model \parencite{Drobyshevskiy.2009.RandomGraphs, Newman.2002}} with $p(\bm{w})$ and variable states with a conditional distribution given the network topology $p(\bm{x} \mid \bm{w})$ in Equation (\ref{eq_10}). Put simply, the probability distribution of the IM can be decomposed into two parts \parencite[see][for details]{Marsman.2023.IdiographicIsing}.
\begin{equation}
    p(\bm{x})_{Ising} =\sum_{\bm{w}} \overbrace{p(\bm{x} \mid \bm{w})}^{\text {graph coloring }} \overbrace{p(\bm{w})}^{\text {random-cluster model} \tag{10} \label{eq_10}}
\end{equation}
 The \textit{Erdős-Rényi model} (ER; \cite{Karonski.1997.RandomGraphs}) models the probability distribution of pairwise edges in a network with $P$ nodes (\ref{eq_10.1}). With distinct edge inclusion probabilities $\theta_{i j}$ for each edge $w_{i j}$. Here $W_{ij}$ is a binary random variable which indicates the presence or absence of an edge in the network and thus describes the probability distribution of different network topologies, where clusters (direct or indirectly connected nodes) are an important concept. Because the ER model is a special case of the \textit{random-cluster model}:
\begin{equation}
    p(\bm{w}) = \frac{1}{Z_R} \overbrace{\prod_{i=1}^{P-1} \prod_{j=i+1}^P \theta_{i j}^{w_{i j}}\left(1-\theta_{i j}\right)^{1-w_{i j}}}^{\text{ER model}} \prod^{\kappa(w)}_{c=1} \lambda_{c} \tag{10.1} \label{eq_10.1}
\end{equation}

With $\kappa(\bm{w)}$ as the number of clusters implied by $\bm{w}$, $\lambda$ a positive cluster weight (favors fewer clusters with $\lambda < 1$, and more clusters with $\lambda > 1$), and $Z_R$ a normalization constant. A random-cluster model assigns different weights to the different clusters, with a unit clustering weight it is an ER model. The main effects of the IM ($\mu_i$) can be related to the cluster weights:
\begin{equation}
    \lambda_c = 2 \cosh \left(\sum_{i \in V_c} \mu_i\right) \geq 2 \tag{10.1.1} \label{eq_10.1.1}
\end{equation}
Where $V_c$ is the set of vertices in cluster $c$. In Graph Theory \parencite{Das.2023.GraphDataScience, Diestel.2025.GraphTheory} node states are usually labeled with colors and the process of assigning states to nodes is then called graph coloring, $p(\bm{x}|\bm{w})$ in Equation (\ref{eq_10}) describes the coloring process of the graph that is induced by $p(\bm{w})$. The graph coloring process assumes that if two variables are connected they must be in the same state and conversely \parencite[see][for details]{Marsman.2023.IdiographicIsing}:
\begin{align}
(W_{ij} = 1) \Rightarrow (X_i = X_j) \notag\\
(X_i \neq X_j) \Rightarrow (W_{ij} = 0) \notag
\end{align}
Since node states align when in the same cluster, cross-sectional correlation of nodes are high when many individuals exhibit the same patterns, therefore individual topologies affect cross-sectional observations. This can be formalized in Equation (\ref{eq_10.1.2}), group-level associations (pairwise interactions $\sigma_{i j}$ in the IM) are related to idiographic edge inclusion probabilities $\theta_{i j}$ in the random-cluster model. 
\begin{equation}
    \sigma_{i j}=-\frac{1}{2} \log \left(1-\theta_{i j}\right) \tag{10.1.2} \label{eq_10.1.2}
\end{equation}
Therefore no associations at the group-level implies no link between two variables in all idiographic networks, but patterns in observed correlations at the group-level can have many different patterns at the individual-level. Therefore homogeneous data is not necessarily required and suggests a reconciliation of the idiographic and nomothetic approaches in network psychometrics \parencite{Marsman.2023.IdiographicIsing}.

Since in practice $Z_R$ in Equation (\ref{eq_10.1}) is difficult to evaluate, other models for the latent topologies can be considered as well, when the ER model is used, the essential connection between idiographic network links and cross-sectional correlations remains, but $p(\bm{x})$ in (\ref{eq_10}) is no longer an IM but a \textit{Divide and Color Model} \parencite{Steif.2019.DivideColor, Marsman.2023.IdiographicIsing}. The encompassing network model can be further extended into the time-series domain by modeling the transition of idiographic topologies into different states with a HMM \parencite{Failenschmid.2023.TimeIsing}.

\subsection{Utilize your Tools (Tutorials)}
So the Psychometric-Toolbox has actually much more to offer than what is currently utilized in empirical research, usually only a small number of statistical models are commonly applied \parencite{Robinaugh.2020.Psychopathology, Punzi.2022.Review, Blanchard.2023.Review, Schumacher.2024.Review}, presumably because these models are accessibly implemented in R-Packages. Furthermore, the analysis of empirical datasets in psychometrics is usually done with only a single model (e.g., \textcite{VillacuraHerrera.2025.Depression-E, Ramm.20255.Depression-E, Curtiss.2022.Depression-E, Groen.2019.Depression-E, Moron.2024.Depression-E, Mullarkey.2019.Depression-E}). But since we presented a large set of statistical models, maybe the default should be applying several models, with different models ideally being able to reveal different patterns, which could enable more robust theoretical interpretations \parencite{Silberzahn.2018}. 

In the following three methods are applied to analyze (Code available at: \url{https://github.com/kekistermann/Exploring-the-PsychometricToolbox}) a psychometric ILD dataset from \textcite{Rowland.2020b.Dataset} that consists of 125 undergraduate students (M age = 22.87; SD = 5.06 years; 77.6\% female) who answered eight items assessing their affective experiences ("happy", "excited", "relaxed", "satisfied", "angry", "anxious", "depressed", "sad") on a visual slider scale from 0 (not at all) to 100 (very much), 6 times a day for 40 days, and were either randomly selected for a weekly mindfulness training (n=61) or a wait-list control condition (n=64).
In \textcite{Rowland.2020b.Dataset}\footnote{Data and code of original analysis: \url{https://osf.io/jmz2n} \parencite[see also][]{Rowland.2016.Dataset, Rowland.2019, Rowland.2020a.Dataset}}, a saturated gVAR model was estimated for each participant to test whether mindfulness training facilitated disengagement from prior affective experiences in daily life, operationalized as changes in temporal affect network density (i.e., the average strength of all possible network connections). However, no significant effect between mindfulness training and temporal affect network density were observed. Examining group differences (control vs. mindfulness intervention) using alternative methods may still be of interest. However, the primary aim here is to illustrate the application of the selected methods rather than to derive new theoretical insights, although it might suggest possibilities (see \href{https://github.com/kekistermann/Exploring-the-PsychometricToolbox}{\textsf{mlVAR\_GIMME.Rmd}} and \href{https://github.com/kekistermann/Exploring-the-PsychometricToolbox}{\textsf{RecurrenceAnalysis.ipynb}} for details).

\subsubsection{1. multilevel VAR (mlVAR)}
\begin{figure*}[!htbp]
    \centering
    \includegraphics[scale=1]{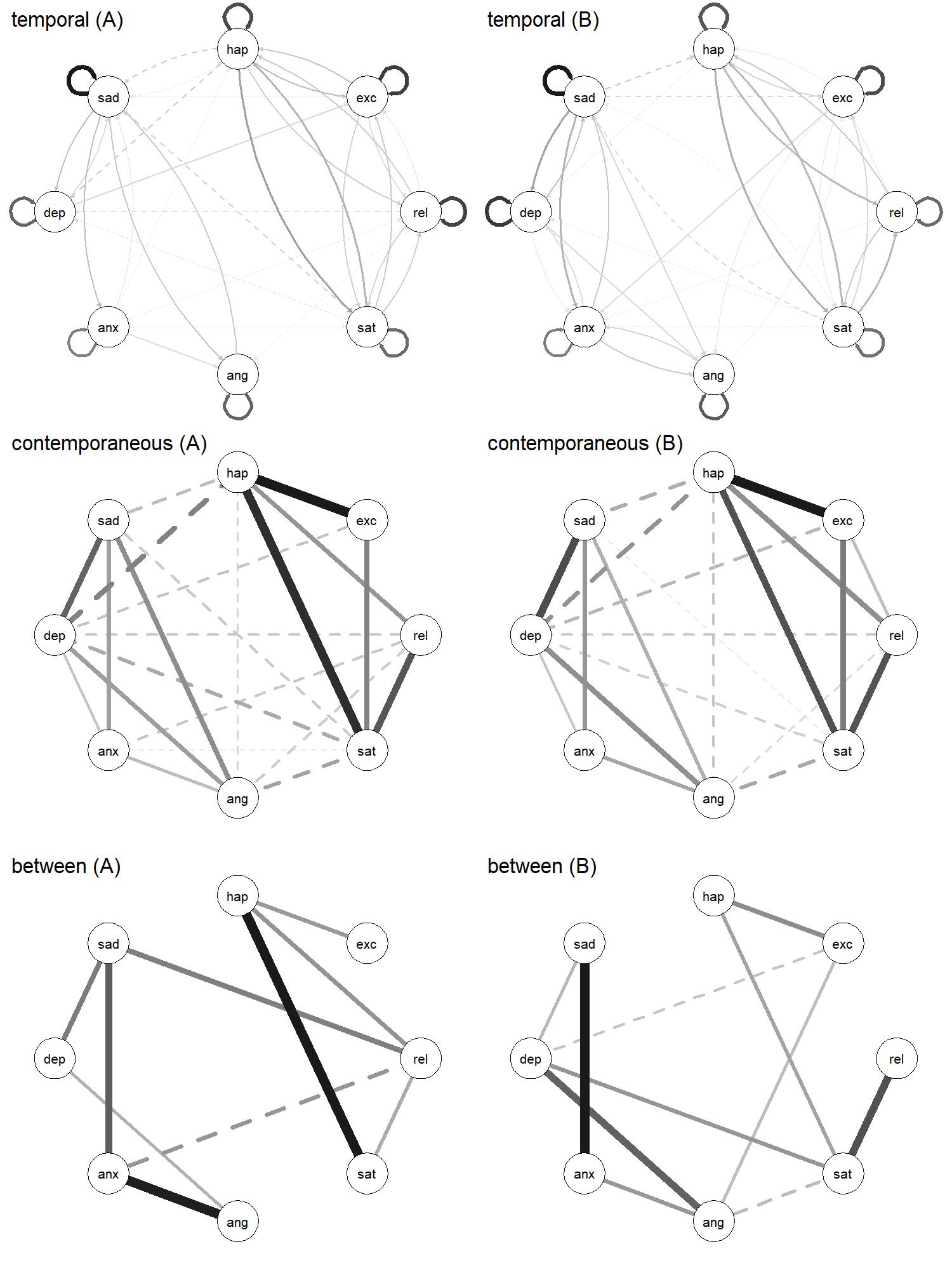}
    \caption{\texttt{mlVAR(...)} output plotted. With control group (A) and intervention group (B). With temporal effects (lag-1 regression weights), contemporaneous relationships (partial correlations) and between-subject effects (partial correlations). Shading and width indicate edge strength, with positive values (solid) and negative values (dashed).}
    \label{fig:mlVAR}
\end{figure*}

The \textit{multilevel VAR} (mlVAR; \cite{Epskamp.2018.GGM-GVAR, Li.2022.mlVAR}) models each person with an individual gVAR model, extending Equation (\ref{eq_6}) to the following:
\begin{align}
\bm{x}_{i, T} \mid \bm{x}_{i, T-1} &= \bm{x}_{i, t-1} \notag\\
& \sim \mathcal{N}\left(\bm{\mu}_i + \bm{B}_i (\bm{x}_{i, t-1} - \bm{\mu}_i), \bm{\Theta}_i\right) \tag{11} \label{eq_11}
\end{align}
where the between-person network is $\bm{K}^{(\Omega)} = \bm{\Omega}^{-1} = var(\bm{\mu})^{-1}$, the person-specific temporal network $\bm{B}_i$, and contemporaneous network $\bm{K}_i^{(\Theta)} = \bm{\Theta}_i^{-1}$. Two estimation methods are implemented in R-Packages, the "pooled and individual LASSO estimation" with \texttt{mlGraphicalVAR(...)} in \href{https://cran.r-project.org/web/packages/graphicalVAR/index.html}{"graphicalVAR"} and used in the original analysis of \textcite{Rowland.2020b.Dataset}, and the "two-step frequentist multilevel" approach which is implemented with \href{https://cran.r-project.org/web/packages/mlVAR/refman/mlVAR.html#mlVAR}{\texttt{mlVAR(...)}} in \href{https://cran.r-project.org/web/packages/mlVAR/index.html}{"mlVAR"} \parencite[see][for details]{Epskamp.2018.GGM-GVAR}. The three matrices can then be visualized as networks in Figure \ref{fig:mlVAR}, where only small differences between the control (A) and mindfulness intervention (B) can be noticed, the between-person networks differ the most. Networks of individuals can be plotted as well, here more pronounced differences in network topology can be spotted (see \href{https://github.com/kekistermann/Exploring-the-PsychometricToolbox}{\textsf{mlVAR\_GIMME.Rmd}}). This suggests that analyzing individual networks, rather than aggregated networks---which is currently the default in empirical research---may offer more valuable insights.

\subsubsection{2. GIMME}
We also estimated the GIMME from Equation (\ref{eq_7}) with the function \href{https://cran.r-project.org/web/packages/gimme/vignettes/gimme_vignette.html}{\texttt{gimme(...)}} in \href{https://cran.r-project.org/web/packages/gimme/index.html}{"gimme"} for each group separately, here again, differences in aggregated networks for the two groups are not that obvious. But when plotting some individual networks, differences in individual network topology seem very apparent (see \href{https://github.com/kekistermann/Exploring-the-PsychometricToolbox}{\textsf{mlVAR\_GIMME.Rmd}}). This presents both an opportunity and a challenge. On the one hand, network visualization offers an intuitive and appealing way of identifying individual differences; on the other, drawing meaningful conclusions from these differences may prove difficult to justify. Figures might also be somewhat misleading because actually numeric differences in edge weights can be very small. Therefore tools and metrics for describing differences in network topologies numerically are needed (e.g., \textcite{Ulitzsch.2023, Hoekstra.2024, Du.2026}).

When setting \texttt{subgroup = TRUE}, we can also estimate the S-GIMME from Equation (\ref{eq_7b}). And this actually does detect two sub-groups, but they do not really correspond to the control/intervention groups (see \href{https://github.com/kekistermann/Exploring-the-PsychometricToolbox}{\textsf{mlVAR\_GIMME.Rmd}}). Therefore, characterizing these two estimated groups in more detail might be interesting to do in future research.

\subsubsection{3. Recurrence Plot}
\begin{figure}[!htbp]
    \centering
    \includegraphics[width=\linewidth]{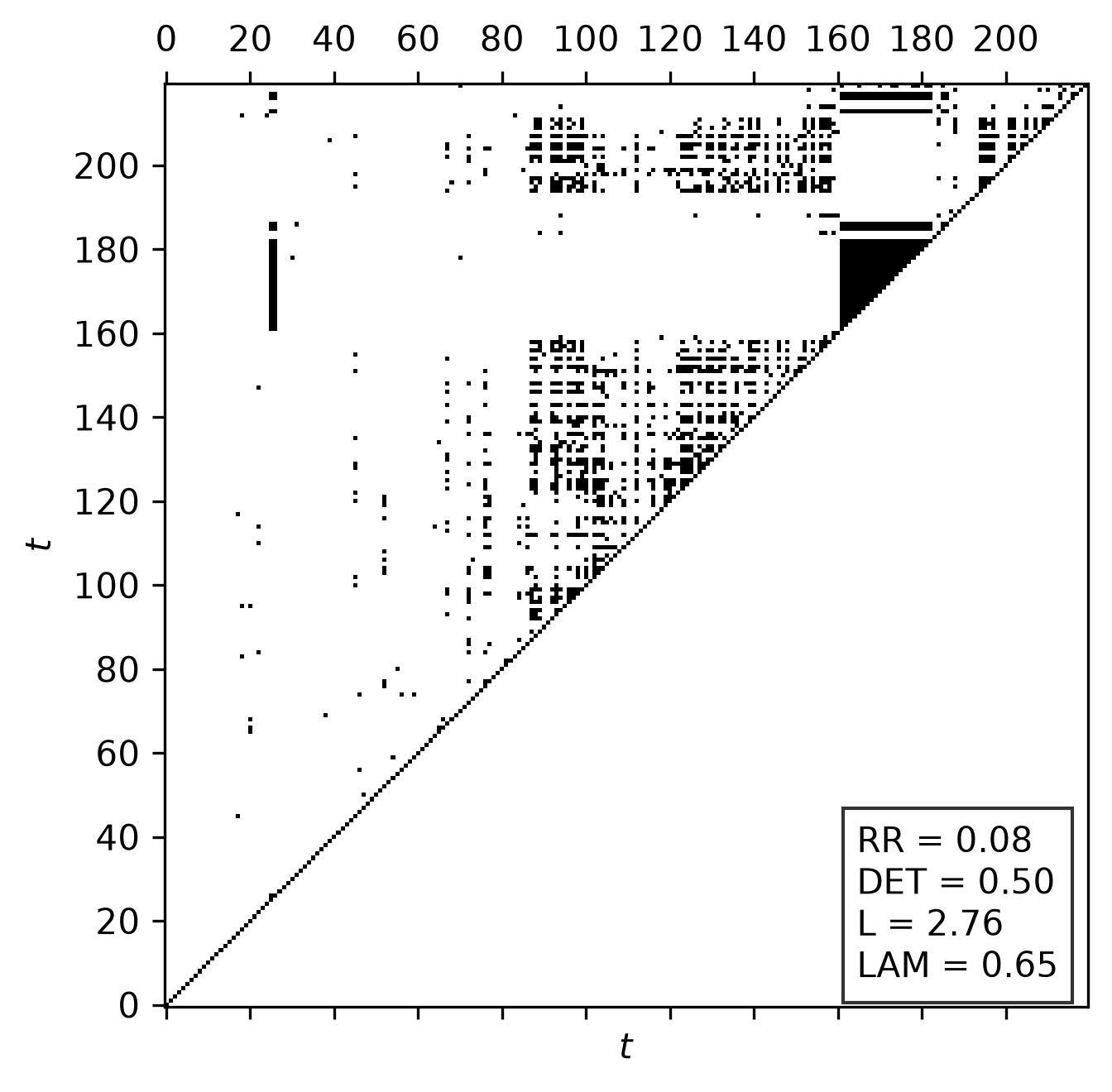}
    \caption{Recurrence Plot of one individual. Metrics: Recurrence Rate (RR), Determinism (DET), average diagonal line Length (L), and Laminarity (LAM).}
    \label{fig:RecurrencePlot_120}
\end{figure}

Since multivariate time-series data in psychometrics likely violates core assumptions for inference with commonly applied statistical models  (e.g., stationarity \parencite{Ryan.2025.Stationarity}), and the practice of interpreting self-reports of human experience analogous to classical physical measurements might be problematic as well \parencite{Hasselman.2020.RN}. The analysis with \textit{Recurrence Plots} (RP; \cite{Marwan.2007.RN, Marwan.2008.RecurrencePlot, Marwan.2023.RecurrenceAnalysis}) is a more foreign but maybe particularly interesting tool for visualizing time series data and its underlying dynamics \parencite{Hirata.2021.RecurrencePlot},  and can be implemented with \texttt{RecurrencePlot(...)} in the \href{https://www.pik-potsdam.de/~donges/pyunicorn}{"pyunicorn"} Python package \parencite{Donges.2015.pyunicorn}. The key concept is recurrence, meaning how the state of a system in some kind of phase/state-space\footnote{Analogously, Kurt Lewin theorized that psychology can be conceptualized as a multidimensional "life space" or psychological field in which behavior and psychological change occur under the influence of multiple interacting factors \parencite{Lewin.1943, Endrejat.2022.KurtLewin, Lewin.1936, Lewin.1938}.} returns over time to past values, in physics this is a fundamental property of dynamical systems and commonly defined as in Equation (\ref{eq_12}).
\begin{equation}
    R_{t,t'}(\epsilon) = \Theta(\epsilon - \left \| \bm{x}_t - \bm{x}_{t'} \right \|), \quad t, t' = 1,..., T \tag{12} \label{eq_12}
\end{equation}
with $T$ the number of time points, $\epsilon$ is a threshold distance, $\Theta(\cdot)$ the Heaviside function and $\|\cdot\|$ a norm (usually Euclidean). Patterns in recurrence (states which are in an $\epsilon$-neighborhood $\Rightarrow \bm{x}_t \approx \bm{x}_{t'}$) can be visualized with a RP as a binary (symmetric) recurrence matrix (see Figure \ref{fig:RecurrencePlot_120}) where
\begin{equation}
    R_{t,t'} = 
        \begin{cases}
            1 & \text{if } \lVert \bm{x}_t - \bm{x}_{t'} \rVert \le \epsilon \\
            0 & \text{otherwise}.
        \end{cases}
    \tag{12.1} \label{eq_12.1}
\end{equation}

How to set the recurrence threshold ($\epsilon$) is crucial, if it is set to small there will be almost no recurrence and if set to big every state is a neighbor of every other state obscuring the dynamics, a common "rule of thumb" is to select it so that the Recurrence Rate (RR) is small \parencite{Marwan.2007.RN}.
The patterns in RPs can be indicators for determinism (diagonal structures), chaos (short diagonal segments due to divergence of trajectories), stochasticity (unstructured recurrence points), and the system remaining in similar states (horizontal/vertical lines). This might be a valuable tool particularly for initial explorations of the data, due to its ease of application and the fact that homogeneity in RP patterns indicates stationarity of the time series. But deriving substantial theoretical explanations from visual interpretation alone is again difficult. 

In \textit{Recurrence Quantification Analysis} (RQA\footnote{With \textit{Recurrence Pattern Correlation} (RPC; \cite{Marghoti.2026.RPC}) being a recent extension which can detect a wider range of recurrence patterns.}; \cite{Marwan.2007.RN}) these patterns can be summarized and quantified. With common metric being the \textit{Recurrence Rate} (RR) as the percentage of recurrence points in the RP, \textit{Determinism} (DET) as the percentage of recurrence points forming diagonal lines, \textit{Average diagonal line length} (L), and \textit{Laminarity} (LAM) as percentage of recurrence points which form vertical lines.

When applied to the dataset (see \href{https://github.com/kekistermann/Exploring-the-PsychometricToolbox}{\textsf{RecurrenceAnalysis.ipynb}}), interesting patterns emerge for some but not all individuals, at least for the same $\epsilon$. In Figure \ref{fig:RecurrencePlot_120} the RP of one participant is shown. The first insight is its non-homogeneity, which strongly indicates non-stationarity of the time-series. The many isolated single points also point to some underlying stochasticity in the process. The presence of some diagonal line structures suggests that segments of the trajectory evolve in a similar manner at times, reflected in the moderate deterministic behavior (DET = 0.50). However, the relatively small average diagonal length (L = 2.76) indicates that this predictability is short-lived, especially when considering that usually there are six observations per day. The noticeable vertical and horizontal line structures and the relatively high Laminarity value (LAM = 0.65) point to phases during which the system remains temporarily in similar states. An interesting question to investigate might be whether the control and intervention groups differ in their RQA metrics. But the theoretical interpretation of these metrics in the psychometric context remains an open question for now.

\section{Statistical Models as Networks} \label{2Section}
As the above should have made clear, purely from the perspective of statistical modeling there is no clear boundary separating "network models" from "classical" psychometric latent variable models. 
But networks also come in different forms, primarily as PGMs \parencite{Maasch.2025.PGM} or Random Graphs in Graph Theory \parencite{Das.2023.GraphDataScience, Diestel.2025.GraphTheory}, yet formal connections already exist \parencite{Farasat.2015, Marsman.2023.IdiographicIsing}. And even though Random Graphs and Graph Theory in general can be seen as the initial basis for the more interdisciplinary effort of \textit{Network Science} \parencite{Hidalgo.2016, Sweet.2025}, they represent fairly independent research fields which have focused on investigating different properties of networks \parencite{Iniguez.2020}. A important distinction is therefore nonetheless between networks where edges are given (e.g., social network, transportation network, citation network) usually encountered in (Social) Network Science and applications where edges first need to be estimated from data (e.g., as partial-correlations in the precision matrix of a GGM) before properties of the network structure can be investigated.
With regards to network psychometrics we can say, at present, this almost exclusively refers to the estimation of statistical pairwise variable interactions with a PGM from psychometric data as some form of covariance\footnote{\textit{S1.1.1 Covariance (cov)} in \parencite{Cliff.2023.SPI}}. But this is not always appropriate \parencite{Slipetz.2024.dcorr}, and a large collection of \textit{Statistics of Pairwise Interactions} (SPIs; \textcite{Cliff.2023.SPI}), including Precision from Equation (\ref{eq_3}), as well as others proposed in psychometrics previously are in principle available here. For example, the \textit{partial distance correlation}\footnote{\textit{S1.2.1 Distance correlation (dcorr)} in \parencite{Cliff.2023.SPI}} which can identify possible non-linear variable relationships \parencite{Slipetz.2024.dcorr}, a extension of the Heller-Heller-Gorfine test\footnote{\textit{S1.2.6 Heller-Heller-Gorfine Independence Criterion (hhg)} in \parencite{Cliff.2023.SPI}} \parencite{Karch.2024.BeyondCorrelation}, and information theoretical measures
in general \parencite{vanSanten.2026}. Perhaps SPIs can be viewed in networks analogous to link functions in GLMs.

\textcite{Marsman.2023.IdiographicIsing} also point out that relations between PGMs and Random Graphs could further our conceptual understanding of psychometric constructs as characterized by individual network topologies. Therefore it might be insightful to establish any regularities in their structure, as it has been done in Network Science where several topological patterns like community structure or small-worldness are commonly observed in many different applications \parencite{Newman.2003.NetworkScience}. Are psychometric topologies similar or different and can we come up with plausible psychological processes that could generate them? 

Thus, these statistical equivalences can also enable conceptual reinterpretations. This can be further illustrated by the relation between \textit{IsingFit} (the eLASSO nodewise L1-regularized logistic-regression approach for estimating the IM \parencite[see][]{Brusco.2019.Ising, vanBorkulo.2014.IsingFit}), \textit{Image Factor Analysis} (IFA; \cite{Guttman.1953}) and IRT outlined in \textcite{Christensen.2021, Borsboom.2023}, where the most central concepts in psychometric are re-framed form a network perspective. \textcite{Borsboom.2023} proposed to conceptualize psychometric constructs as pragmatically selected sets of \textit{structurally connected}\footnote{$\underleftrightarrow{?}$ "Morinian Interaction" \parencite[in][]{Estrada.2024.ComplexSystem}} indicators (variables/items) of psychological phenomena. In that light \textit{Unidimensionality} is approximated by a densely connected network, which is statistically equivalent to a unidimensional factor model (e.g., Curie-Weiss model $\Leftrightarrow$ Rasch model \parencite{Marsman.2018.IRT-Ising}) and can therefore be seen as a measure of network homogeneity. Furthermore, mean-field approximations of PGMs \parencite{vanderMaas.2024} can be viewed as approximating the first factor in IFA. \textit{Reliablity} in a network context reflects how strongly variables are connected, which indicates the degree to which variable states are predictable from their neighbors, strong connections represent a low-entropy system in which variable values align. For \textit{Validity} the network perspective implies that psychometric constructs (e.g., psychopathologies, intelligence, personality) don't refer to latent variables that cause item responses, but that the variables assessed through item responses are in fact part of a "construct network". Within this framework, a psychometric construct can be viewed as a collection of \textit{structurally connected} variables that interact with one another, regularities in network topologies could then correspond to psychological (phenomenological) states. And these interactions, observed empirically as correlations between items, can be 
 \begin{quote}
    approximated by pairwise interactions; these pairwise interactions, in turn, generate probability distributions that tend to fit latent variable models reasonably well \parencite{Borsboom.2023}.
\end{quote}

\section{Networks as Conceptual Framework} \label{3Section}
After the initial relations between statistical models in psychometrics were established, the importance of differentiating between empirical/methodological and conceptual/theoretical problems has been stressed \parencite{Bringmann.2018.Psychopathology, vanBork.2021}. But without any apparent cumulative theory building (e.g., in research on depression \parencite[see][]{VillacuraHerrera.2025.Depression-E, Ramm.20255.Depression-E, Cai.2024.Depression-E, Curtiss.2022.Depression-E, Groen.2019.Depression-E, Moron.2024.Depression-E, Mullarkey.2019.Depression-E}), the empirical literature of network psychometrics is nevertheless steadily growing \parencite{Robinaugh.2020.Psychopathology}. The conflation of statistical and theoretical models due to the lack of strong formal theories has been identified as a fundamental shortcoming of psychology \parencite{Fried.2020.Statistical-Conceptual, Kerschke.2025}. However there is progress made in attempts to mitigate this issue, with a general discussion and outline of a methodology for productive explanations in psychology \parencite{vanDongen.2024}, awareness of how conceptual progress is essential and complementary in addressing empirical problems \parencite{Adolfi.2024, Grimm.2026}, as well as the development of general theory construction methodologies \parencite{HawkinsElder.2020.TheoryConstruction, Borsboom.2021.TheoryConstruction, Robinaugh.2021.TheoryConstruction, Haslbeck.2022.TheoryConstruction}. Computational models could be essential in this research effort \parencite{Guest.2021, Evers.2026.PsyTheory}, and represent a valuable tool for constructing formal theories when empirical evidence is spares or indirect \parencite{Woensdregt.2024}.

\paragraph{Idiographic Psychology}
Furthermore, the vast majority of empirical research in network psychometrics analyze cross-sectional data \parencite{Robinaugh.2020.Psychopathology, Punzi.2022.Review, Schumacher.2024.Review}, but interpreting such finding as informative of within-person processes has been recognized as an ecological fallacy \parencite{Hamaker.2025.ILD}, the inability of group-to-individual generalizability is a serious issue \parencite{Fisher.2018, Speelman.2020, Richters.2021, Hunter.2024}. The nature of ILD allows us---at least in principle---to differentiate between-person (nomothetic) and within-person (idiographic) patterns (cross-sectional data = a mix of between- and within-person patterns), as well as learning about between-person differences in within-person dynamics \parencite{Hamaker.2025.ILD, Borsboom.2024.IdiographicPsy}. The first methodological bridge for differentiating between individual- and group-level patterns has been introduced earlier with the \textit{Idiographic IM} \parencite{Marsman.2023.IdiographicIsing}.

\paragraph{Graph Theory}
The goal should therefore be to utilize mathematical formalisms that can adequately represent the relevant characteristics of psychometric constructs and in doing so connect psychological theory to empirical observations \parencite{Borsboom.2023}. Graph Theory represents an extensive researched formalism that can be applied to investigate properties of complex dynamical systems as networks \parencite{Porter.2016}, and could potentially be utilized more in psychometrics for doing that. But it is important to recognize that this formalism is not assumption free either, but keeping this in mind, they can be chosen appropriately to match the applied context \parencite{ZamoraLopez.2024}. Some vague suggestion for the uses of Graph Theory in the social and behavioral sciences have been made some time ago already and include the potential for better quantifying structural and qualitative aspects of the constructs of interest \parencite{Harary.1960.GraphTheroy-SocialScience, Tapiero.1972.GraphTheroy-BehavioralScience}. But this type of research is in its infancy and a more detailed and perhaps structured investigation of how graph-theoretic concepts could add to the psychometric-toolbox should probably be conducted in the future.

\paragraph{Topological Explanations} 
Most generally this opens up a interesting connection to a discussion in the philosophy of science about \textit{mechanistic vs. topological} explanations \parencite{Kostic.2018a, Kostic.2018b, Kostic.2020, Kostic.2021, Kostic.2023, Kostic.2025, Huneman.2025}.
\begin{quote}
    Topological explanations explain the dynamics of complex systems by making use of topological properties, i.e., properties of a complex system that are mathematically quantified using graph theory \parencite{Boer.2021}
\end{quote}
, with the appealing feature of being able to provide explanations of non-decomposable systems \parencite{Rathkopf.2018}. In \textcite{Skowron.2021} topology is viewed as a potential unifying framework for diverse approaches, specifically phenomenological \parencite{Prentner.2025} and neuroscientific \parencite{Sizemore.2019.TDA} ones, but also in other domains \parencite{Skowron.2023}.

\section{Conclusion}
As we have tried to illustrate, the psychometric-toolbox has actually much more to offer then what is currently utilized and there are many more interesting paths forward. However, the diagram of model relations presented here has certain limitations. While it provides a more convenient overview and serves as an accessible entry point for more detailed exploration, retrieving the necessary details for each model remains tedious. Therefore, more systematic formalizations should be developed in subsequent work. Furthermore, since the development of statistical methodology is continuously evolving, documenting available statistical models and their relations should ideally be organized in a format that allows for continuous updating and editing, ideally as an collective effort of the research community. And lastly, methodological developments needs to be linked more explicitly and systematically to the processes of theory construction in psychology, to the benefit of both.

\section{Supplementary Materials --- Extending the Psychometric-Toolbox}

\subsection{PGM, KIM, VDAR, GRG, (T)ERGM}
Let's start by considering PGMs\footnote{the term 'Probabilistic Network Models (PNMs)' is also used in psychometrics \parencite{Finnemann.2026.Potts} and other fields (e.g., Climatology \parencite{Graafland.2020}, Bioinformatics \parencite{Graafland.2022}), but since PGMs are an established term in Machine Learning for such models, maybe we should reserve the term PNMs more generally for models with aspects of both Random Graph Models and PGMs, but again, the boundary is fuzzy here as well \parencite{Farasat.2015, Marsman.2023.IdiographicIsing}.} again and assume that variable interactions result in the coordination of variable states, six such base models for different variable types are presented in \textcite{Finnemann.2026.Potts} and include the IM for binary variables. The IM has also been extended in econometrics into the time-series domain with \textit{Multiple Time Series IM} \parencite{Takaishi.2015, Takaishi.2016} and in physics with the \textit{Kinetic IM} (KIM; \cite{Aguilera.2021.KIM}) which has been formally related to the \textit{Vector Discrete AR} (VDAR; \cite{Campajola.2021.KIM/VDAR}) model, the discrete analogue to the well-known VAR model\footnote{$\Rightarrow$ IM (binary MRF) $\rightarrow$ KIM $\Longleftrightarrow$ VDAR --- VAR $\rightarrow$ gVAR $\leftarrow$ GGM (gaussian MRF); Autoregressive and state-space models are GMRFs \parencite[see][]{Rue.2005.GMRF}.}. Furthermore, the ER model has been generalized with \textit{Generalized Random Graph} (GRG\footnote{Temporally Generalized Random Graphs (TGRG; \cite{Mazzarisi.2020.TGRG}), Discrete Auto Regressive Temporally Generalized Random Graphs (DAR-TGRG; \cite{Mazzarisi.2020.TGRG})}; \cite{Shi.2009.GRG}) and extended in the \textit{Exponential Random Graph Model} (ERGM\footnote{Related to GLMs \parencite{Kolaczyk.2020} and can be re-defined as a PGM \parencite{Farasat.2015}.}; \cite{Zhang.2024.ERGM}), which has with the \textit{Temporal ERGM} (TERGM; \cite{Hanneke.2010.TERGM}) a time-series extension as well and can also be related back to the KIM \parencite{Campajola.2022.KIM-TERGM}.

\setcounter{figure}{7}
\begin{figure}[h]
\centering
\begin{tikzpicture}[
    font=\small,
    model/.style={
        rectangle,
        thick,
        minimum width = 0.1cm,
        minimum height = 0.1cm,
        align = center,
        inner sep=1pt
    },
    binary-categorical/.style={
        draw,
        rectangle,
        dashed, ultra thick,
        minimum width = 0.1cm,
        minimum height = 0.1cm,
        align = center,
        inner sep=1pt
    },    
    physics_cross/.style={model, fill=blue!10},
    physics_time/.style={model, fill=blue!45},
    ML_cross/.style={model, fill=cyan!10},
    ML_time/.style={model, fill=cyan!45},
    Graph_cross/.style={model, fill=magenta!15},
    Graph_time/.style={model, fill=magenta!45},
    psychometrics_cross/.style={model, fill=green!10},
    psychometrics_time/.style={model, fill=green!45},
    network_cross/.style={model, fill=teal!10},
    network_time/.style={model, fill=teal!45},
    econometrics/.style={model, fill=orange!45},
    relation/.style={-{Latex[length=2mm]}, thick}
]
\node[physics_cross] (IM) at (-6,1.5) {
\textbf{Ising Model (IM)}\\[1pt]
$p(\bm{x}) \propto \exp\!\left( \sum_i^{n} \mu_i x_i + \sum_{<i,j>} \sigma_{ij} x_i x_j\right)$
};

\node[physics_time] (KIM) at (-0.5,2.5) {
\textbf{Kinectic IM}
};

\node[econometrics] (VDAR) at (2.5,2.5) {
\textbf{VDAR}
};

\node[Graph_time] (TERGM) at (2.5,3.3) {
\textbf{TERGM}
};

\node[network_cross] (IdiographicIM) at (-6,4) {
\textbf{Idiographic IM}\\[1pt]
$p(\bm{x}) = \sum_{\bm{w}} \overbrace{p(\bm{x} \mid \bm{w})}^{\text {graph coloring }} p(\bm{w})$
};

\node[network_cross] (DCM) at (-1,4) {
\textbf{Divide \& Color}\\[1pt]
$p(\bm{x}) = \sum_{\bm{w}} \overbrace{p(\bm{x} \mid \bm{w})}^{\text {graph coloring }} p(\bm{w})$
};

\node[Graph_cross] (ERGM) at (2.5,5) {
\textbf{ERGM}
};

\node[ML_cross] (GLM) at (4,4) {
\textbf{GLM}
};

\node[psychometrics_cross] (SEM) at (4,2.5) {
\textbf{SEM}
};

\node[psychometrics_time] (MSEM-RS) at (4,1.5) {
\textbf{MSEM-RS}
};

\node[physics_time] (cusp) at (1,1.5) {
\textbf{Cusp Catastrophe}
};

\node[Graph_cross] (ER) at (2,6.35) {
\textbf{Erd\H{o}s--R\'enyi}
};

\node[Graph_cross] (RCM) at (-5,6.35) {
\textbf{Random Cluster Model}\\[1pt]
$p(\bm{w}) \propto \prod_{i=1}^{n-1} \prod_{j=i+1}^n \theta_{i j}^{w_{i j}}\left(1-\theta_{i j}\right)^{1-w_{i j}} \lambda^{\kappa(\bm{w})}$
};

\draw[relation] (IM.east) -- (KIM.west) node[fill=white, midway] {$T>1$};
\draw[relation] (ER) -- (DCM) node[fill=white, midway] {$p(\bm{w})$};
\draw[relation] (RCM) -- (IdiographicIM) node[fill=white, midway] {$p(\bm{w})$};
\draw[relation] (RCM) -- (ER) node[fill=white, midway] {$\lambda^{\kappa(\bm{w})}=1$};
\draw[relation] (ERGM) -- (TERGM) node[fill=white, midway] {$T > 1$};
\draw[] (IM) -- (IdiographicIM.south) node[fill=white, midway] {$\sigma_{i j} = -\frac{1}{2} \log \left(1-\theta_{i j}\right)$};
\draw[relation] (ERGM) -- (TERGM) node[fill=white, midway] {$T > 1$};
\draw[] (TERGM.west) -- (KIM.east);
\draw[] (VDAR) -- (KIM);
\draw[relation] (ERGM) -- (ER);
\draw[relation] (RCM) -- (ER) node[fill=white, midway] {$\lambda^{\kappa(\bm{w})}=1$};
\draw[] (ERGM) -- (GLM);
\draw[relation] (SEM) -- (GLM);
\draw[relation] (SEM) -- (MSEM-RS);
\draw[dashed] (MSEM-RS) -- (cusp);
\draw[dashed] (IM.east) -- (cusp.west) node[fill=white, midway, font=\small] {mean-field};

\end{tikzpicture}
\caption{Erd\H{o}s--R\'enyi (ER or Random Graph Model), Exponential Random Graph Model (ERGM), Temporal Exponential Random Graph Model (TERGM), Vector Discrete Autoregressive (VDAR), Mixture SEM with Regime-Switching (MSEM-RS). \underline{Areas of Research}: Psychometrics (\colorbox{green!10}{cross-sectional}/ / \colorbox{green!45}{time-series}), \colorbox{teal!10}{Network Psychometrics}, \colorbox{orange!45}{Econometrics}, \colorbox{cyan!10}{Statistics \& ML}, Graph Theory \& Network Science (\colorbox{magenta!10}{cross-sectional}/ \colorbox{magenta!45}{time-series}), Physics (\colorbox{blue!10}{cross-sectional} / \colorbox{blue!45}{time-series})}
\label{fig:ModelDiagram_4}
\end{figure}

\subsection{PGMs --- Random Graph Models (Graph Theory)}
The Idiographic IM presents an interesting formal relation between PGMs and Random Graph models from Graph Theory usually applied in \textbf{Network Science} \parencite{Newman.2003.NetworkScience, Coscia.2025}. These are research fields that have to a large extent developed independent from each other, which makes it sometimes unclear how they relate \parencite{Hidalgo.2016, Iniguez.2020, Sweet.2025}. This is an issue illustrated by the common use of node centrality measures for quantifying relative importance of nodes in network psychometrics \parencite{Borsboom.2021.NetworkPsychometrics}, which has been criticized because these centrality measures were initially proposed for (unweighted) random graph models \parencite{Bringmann.2019.Centrality, Spiller.2020.Centrality, Hallquist.2021.Centraility}. But they could conceivably be applied to weighted networks as well \parencite{Opsahl.2010.Centrality} and due to this relation therefore also to PGMs such as the IM or GGM.

Graph Theory \parencite{Das.2023.GraphDataScience, Diestel.2025.GraphTheory} is a branch of mathematics that studies graphs, a mathematical structures for modeling (pairwise) relationships between objects. A graph $G$ is typically defined as an ordered pair $G = (V, E)$ where $V$ is a set of vertices (or nodes/variables), and $E \subseteq \{(i, j) \mid i, j \in V\}$ is a set of edges, which are pairs of vertices representing connections between them. A common distinction we have already encountered previously is between directed graphs where $E$ consists of ordered pairs of vertices, edges have a direction ($(i, j) \neq (j, i)$ $\rightarrow$ e.g., DAG/BN) and undirected graphs where $E$ consists of unordered pairs of vertices, edges have no direction ($(u, v) = (v, u)$ $\rightarrow$ e.g., IM, GGM). Both can either have weighted edges, meaning each edge $e \in E$ is associated with a weight or strength (e.g, partial-correlation coefficient) or unweighted edges (e.g., ER model). And as previously mentioned, investigating the characteristics of graphs representing psychometrics constructs has been independently proposed again more recently in \textcite{Marsman.2023.IdiographicIsing}. With some specific application being the use of a graph-theoretic similarity measure for comparing weighted adjacency matrices \parencite{Ulitzsch.2023}, and \textit{Perturbation Graphs} for assessing experimental interventions and possibly discovering causal relations in psychology \parencite{Waldorp.2025.Perturbation-Graphs}. Symmetry properties of graph limits and their implications for studying invariant sub-spaces in dynamical systems are investigated in \textcite{Bick.2025}, which could conceivably provide null-models for correlation networks \parencite{Masuda.2025}. A interesting discussion about learning Graph structure from time-series data and the duality between predictability and reconstructability is given in \textcite{Murphy.2024}\footnote{$\underleftrightarrow{?}$ 'Weak vs. Strong Complexity' \parencite{Hasselman.2022.EWS} \& 'nearly decomposable system vs. non-decomposable system' \parencite{Rathkopf.2018, Boer.2021}}. Furthermore, in \textcite{Christensen.2023.wTO} a statistic called \textit{weighted topological overlap} (wTO) is presented which quantifies the extent to which two variables have similar partial correlations to other variables in a sparse matrix, \cite{Ulitzsch.2023} present a graph-theoretic similarity measure for comparing weighted adjacency matrices, and in \cite{Sjobeck.2024} the \textit{Pairwise Approximate Spatiotemporal Symmetry} (PASS) algorithm can identify symmetries in pairwise time-series data.

\subsection{IM --- Cusp --- MSEM-RS}
Another interesting connection here might be helpful in addressing the issue of interpretability with increasing model complexity (number of parameters) and can potentially help in enabling advances in research on some conceptual issues in psychology \parencite{vanderMaas.2024}. In Catastrophe Theory\footnote{a branch of Bifurcation Theory \parencite{Kuznetsov.2023.BifurcationTheory} and applied in the study of dynamical systems in \textbf{Dynamic Systems Theory} (DST; \cite{Connell.2017.DST, Olthof.2023.DST-Psychopathology})} \parencite{Montaldi.2021.CatastropheTheory, Stewart.1983.CatastropheTheoryPsych}, continuous and gradual changes in one or more control parameters of a complex system can lead to sudden and discontinuous changes in its dynamics. The Cusp \parencite{MabireYon.2026.Cusp} is a specific type of catastrophe model with two control parameters, a bifurcation parameter (which can cause instability) and an asymmetry parameter (stable linear relationship), and has been applied in psychology to explain characteristics of depression \parencite{Cramer.2016.Depression}, visual perception \parencite{Ploeger.2002} and attitude change \parencite{vanderMaas.2003}. Cusp dynamics have also been associated with the Ising model in network psychometrics \parencite{Loossens.2020.Affect, Finnemann.2021.Ising, vanderMaas.2017.Intelligence, Cramer.2016.Depression}, and in the case of a fully connected IM (also called the Curie—Weiss model), with its mean-field behavior \parencite{vanderMaas.2024}. Nevertheless, its practical use has been constrained by difficulties in performing model fitting and model comparison, to address this issue a modeling framework utilizing Mixture Structural Equation Models (MSEM) for cross-sectional data, or MSEM with regime-switching (MSEM-RS) for panel-data has been proposed in \textcite{Chow.2015.MSEM-RS}.

\subsection{Complex Dynamical Systems - Perspective}
When trying to model psychological processes like individual developmental pathways the adoption of a complex dynamical systems perspective can be very beneficial \parencite{Hasselman.2023}, and has been invoked in psychometrics for re-conceptualizing depression \parencite{Cramer.2016.Depression}, psychopathology in general \parencite{Olthof.2023.DST-Psychopathology}, change-processes in psychotherapy \parencite{Hayes.2020.Psychotherapy}, substance use disorder \parencite{Witkiewitz.2025.SubstanceUseDisorder}, transitions in mood \parencite{Waldorp.2020.Mood}, phenomenology in psychopathology \parencite{Kyzar.2025.Phenomenology}, as well as in educational psychology \parencite{Hilpert.2018.Education}. This also includes utilizing tool like \textit{Causal Loop Diagrams} (CLDs\footnote{e.g., of Depression \parencite{Wittenborn.2016.CLD-Depression}}; \cite{BarbrookJohnson.2022.CLD, Crielaard.2024.aCLD}) and \textit{Perceived Causal Networks} (PECAN; \cite{Vogel.2025.PCN, Burger.2024.PCN, Scholten.2025.PCN}) which can be combined with VAR models \parencite{Burger.2022.PREMISE}.

A clarifying discussion of what is meant by "Complex (Dynamical) System" can be found in \textcite{Estrada.2024.ComplexSystem} and \textcite{Mazzocchi.2025.ComplexSystem}. For an elaboration of how this could be applied in psychology see \textcite{vanderMaas.2024}. Patterns in ILD are likely complex, because EMA data is characterized by complex long-range temporal dependencies \parencite{Olthof.2020, Olthof.2024}, and psychological self-ratings exhibit complex dynamics as well \parencite{Olthof.2020.SelfRatings, Olthof.2023, Leising.2025}. Different metrics can be used for quantifying the complexity in time-series data, and \textcite{George.2025} investigated how they relate to dynamics. Furthermore, this can be tied back very interestingly to statistical models with \textit{Coevolving Latent Space Networks with Attractors} (CLSNA; \cite{Pan.2025.CLSNA}).

\subsubsection{Early Warning Signals}
A concrete example of how this this line of research could become relevant in clinical psychology are \textit{Early Warning Signals} (EWS; \cite{Evers.2024.EWS}). A particular EWS is \textit{Critical Slowing Down} (CSD) which is an indication of bifurcations \parencite{Evers.2024.EWS}, and has been applied in psychology to assess how this is related to changes in depression symptoms \parencite{Curtiss.2023.EWS, Fried.2023.EWS} and mood disorder during psychotherapy \parencite{Olthof.2020.EWS}. Specifically, in RNs the geometry of the phase space can function as a EWS \parencite{Hasselman.2022.EWS}. But more research is needed for the appropriate application of EWS in clinical practice \parencite{Cui.2025.EWS}, current modeling approaches are limited, and theoretical foundations of psychological phenomena have to evolve with methods for data analysis in tandem here as well \parencite{Helmich.2024.EWS}. EWS are also researched in physiology (e.g., \parencite{Legault.2024.EWS}), where networks in the context of complex dynamical systems are investigated as well \parencite{Scholl.2022}.

\subsection{Stochastic Block Model}
It has also been proposed in psychometrics to combine the \textit{Stochastic Block Model} (SBM\footnote{for relations to other models \parencite[see][]{Kim.2018}}; \cite{Holland.1983.SBM, Peixoto.2019.SBM}) with the IM to account for clustering of variables \parencite{Sekulovski.2026.OMRF-SBM}, in other research areas the SBM and IM have been combined as well \parencite{Berthet.2019.Ising-SBM, Ye.2020.Ising-SBM, Zhao.2021.Ising-SBM, Zhang.2023.Ising-SBM}.

\subsection{ARMA, DLM, ART}
Additionally to the already considered VAR\footnote{Variants for ILD: ALS VAR \parencite{Bulteel.2016.ALS-VAR}, scGVAR \parencite{Park.2024.sVAR, Park.2024.scGVAR}, multi-VAR framework \parencite{Fisher.2022.multi-VAR, Fisher.2024.multi-VAR, Crawford.2026.multi-VAR}, mlVAR \parencite{Li.2022.mlVAR, Bringmann.2013.mlVAR, Epskamp.2018.GGM-GVAR}} and VDAR model, the ARMA model family is actually much bigger \parencite{Holan.2010.ARMA}. The Threshold AR (TAR; \cite{Tong.1980.TAR}) has been incorporated into psychometrics specifically for modeling hysteresis in psychological processes (HysTAR; \cite{Jong.2023.HysTAR/MSEM-RS}). More generally again, ARMA models can also be formulated as \textit{Dynamic Linear Models} (DLMs; \cite{Laine.2020.DLM, Carvalho.2007.Bayesian-DLM}), and perhaps interesting for ILD applications is the \textit{Autoregressive Tensor}\footnote{Introduction to Tensors \parencite{Bi.2021.Tensor, Auddy.2025.Tensor}. The IM can also be extended to allow for modeling higher-order dependencies with the \textit{p-Tensor IM} \parencite{Mukherjee.2022.TensorIsing}} model (ART; \cite{Billio.2023.ART}) which can include a HMM for modeling structural changes \parencite{Billio.2024.ART}.

\subsection{EGA, EGM, MNM, MAR}
In \textit{Exploratory Graph Analysis} (EGA; \cite{Golino.2017.EGA}) a clustering algorithm is applied to a GGM, this has also been extended to \textit{Dynamic EGA} (DynEGA; \cite{Golino.2022.DynEGA}) and generalized with the \textit{Exploratory Graph Model} (EGM; \cite{Christensen.2024.EGM}). The GGM has also been extended to include moderation effects in the \textit{Moderation Network Model} (MNM; \cite{Haslbeck.2021.MNM}), both the GGM and MNM are special cases of the \textit{k-order Mixed Graphical Model} (MGM; \cite{Yang.2014.MGM, Sedgewick.2016.MGM, Haslbeck.2020.k-orderMGM}). And \textit{Moderated AR} (MAR; \cite{Ernst.2024.MAR}) models exist as well.

\setcounter{figure}{10}
\begin{figure}[H]
\centering
\begin{tikzpicture}[
    font=\small,
    model/.style={
        rectangle,
        thick,
        minimum width = 0.1cm,
        minimum height = 0.1cm,
        align = center,
        inner sep=1pt
    },
    binary-categorical/.style={
        draw,
        rectangle,
        dashed, ultra thick,
        minimum width = 0.1cm,
        minimum height = 0.1cm,
        align = center,
        inner sep=1pt
    },    
    physics_cross/.style={model, fill=blue!10},
    physics_time/.style={model, fill=blue!45},
    ML_cross/.style={model, fill=cyan!10},
    ML_time/.style={model, fill=cyan!45},
    Graph_cross/.style={model, fill=magenta!15},
    Graph_time/.style={model, fill=magenta!45},
    psychometrics_cross/.style={model, fill=green!10},
    psychometrics_time/.style={model, fill=green!45},
    network_cross/.style={model, fill=teal!10},
    network_time/.style={model, fill=teal!45},
    econometrics/.style={model, fill=orange!45},
    relation/.style={-{Latex[length=2mm]}, thick}
]
\node[ML_cross] (GGM) at (-2,-1) {
\textbf{GGM}
};

\node[physics_time] (RP/RN) at (-3,1) {
\textbf{RP $\rightarrow$ RN}
};

\node[physics_cross] (IM) at (-11,1) {
\textbf{Ising Model}
};

\node[ML_cross] (MRF) at (-13,-1) {
\textbf{MRF}
};

\node[ML_cross] (ETM) at (-7,1) {
\textbf{Exponential Trace Model}
};

\node[network_cross] (MNM) at (-5.5,0) {
\textbf{MNM}
};

\node[ML_cross] (MGM) at (-8.5,0) {
\textbf{MGM}
};

\node[ML_cross] (BN) at (-13,-3) {
\textbf{(D)BN}
};

\node[ML_cross] (HRF) at (-11,-3) {
\textbf{HRF}
};

\node[ML_time] (DHRF) at (-8,-3) {
\textbf{D-HRF}
};

\node[ML_time] (TD-MRF) at (-8,-2) {
\textbf{TD-MRF}\\[4pt]
$p(\bm{x}) \propto \prod_c \psi_c(\bm{x}_{c}^{t})$
};

\node[ML_time] (HMM) at (-8.5,-4) {
\textbf{Hidden Markov Model (HMM)}
};

\node[ML_time] (DynamicGGM) at (-5,-2.5) {
\textbf{Dynamic GGM}\\[4pt]
$\bm{\Sigma}^t = \left(\bm{K}^t\right)^{-1}$
};

\node[ML_time] (TVGL) at (-3.5,-5) {
\textbf{TVGL}
};

\node[ML_time] (LTGL) at (-6.5,-5) {
\textbf{(L)TGL$_{\kappa/p}$}
};

\node[ML_time] (TAGM) at (-4.5,-4) {
\textbf{TAGM}
};

\node[ML_cross] (LVGGM) at (0.5,-2.5) {
\textbf{LVGGM}
};

\node[ML_cross] (GGIM) at (0,-1) {
\textbf{GGIM}
};

\node[ML_cross] (cdexGGM) at (-0.5,-4.5) {
\textbf{cdexGGM}
};

\node[ML_cross] (latentGGM) at (-2,-5.5) {
\textbf{latent GGM}
};

\node[ML_cross] (RandomGGM) at (0,0.5) {
\textbf{Random GGM}
};

\draw[dashed] (GGM) -- (RP/RN);
\draw[relation] (ETM.east) -- (GGM);
\draw[relation] (ETM) -- (IM.east);
\draw[relation] (MRF) -- (IM) node[fill=white, midway, font=\small] {undirected \& binary};
\draw[relation] (MRF) -- (GGM) node[fill=white, midway, font=\small, sloped] {undirected \& Gaussian};
\draw[relation] (GGM) -- (MNM);
\draw[relation] (MGM) -- (MNM);
\draw[relation] (MGM) -- (MRF);
\draw[relation] (BN.south) -- (HMM.west);
\draw[relation] (HMM) -- (TAGM);
\draw[relation] (GGM) -- (TAGM.north east);
\draw[relation] (GGM) -- (TVGL) node[fill=white, sloped, near end] {$T > 1$};
\draw[relation] (GGM) -- (DynamicGGM.north) node[fill=white, midway] {$T>1$};
\draw[relation] (TVGL) -- (LTGL) node[fill=white, midway] {$\bm{L}$};
\draw[relation] (MRF) -- (BN) node[fill=white, midway, font=\small] {DAG};
\draw[relation] (TD-MRF.west) -- (MRF) node[fill=white, midway, sloped] {$T=1$};
\draw[relation] (DHRF) -- (HRF) node[fill=white, midway] {$T=1$};
\draw[relation] (BN) -- (HRF);
\draw[relation] (MRF) -- (HRF);
\draw[relation] (RandomGGM) -- (GGM) node[fill=white, midway, sloped] {$a_{ij} =1$};
\draw[relation] (cdexGGM) -- (GGM) node[fill=white, midway, sloped] {$x_m^{(h)} = 0 \: \forall \: h$};
\draw[relation] (GGM) -- (LVGGM.west) node[fill=white, midway, sloped] {$- \bm{L}_{\bm{\Omega}}$};
\draw[relation] (GGM) -- (GGIM);
\draw[relation] (GGM) -- (latentGGM) node[fill=white, midway, sloped, font=\small] {binary/mixed};

\end{tikzpicture}
\caption{Markov Random Field (MRF), Time-Dynamic MRF (TD-MRF), Dynamic Bayesian Network (DBN), Hybrid Random Field (HRF), Dynamic HRF (D-HRF), Mixed Graphical Model (MGM), Moderated Network Model (MNM), Gaussian Graphical Model (GGM), Gaussian Graphical Interaction Model (GGIM), Latent Variable GGM (LVGGM),  ovariate-dependent GGM (cdexGGM), Time Adaptive Gaussian Model (TAGM), Time-Varying
Graphical Lasso (TVGL), Latent Variable Time-Varying Graphical Lasso
(LTGL). Colors: \colorbox{teal!10}{Network Psychometrics}, Statistics \& ML (\colorbox{cyan!10}{cross-sectional} / \colorbox{cyan!45}{time-series}), Physics (\colorbox{blue!10}{cross-sectional}/ / \colorbox{blue!45}{time-series})}
\label{fig:ModelDiagram_5}
\end{figure}

\subsection{GGMs in ML (Figure \ref{fig:ModelDiagram_5})}
In the Machine Learning (ML) literature many more variants of the GGM have been proposed (\textit{Latent Variable GGM} (LVGGM; \cite{Wang.2023.PCA-LVGGM, Li.2023.LVGGM}), \textit{latent GGM} \parencite{Zheng.2022.LatentGGM}, \textit{covariate-dependent GGM} (cdexGGM; \cite{Wang.2025.cdexGGM}), \textit{Random GGM} \parencite{Cai.2017.RandomGGM}, \textit{Gaussian Graphical Interaction Model} (GGIM; \cite{Fitch.2019.GGIM})), with time-series extensions existing as well (\textit{Time-Varying Graphical Lasso} (TVGL; \cite{Masuda.2016, Hallac.2017.TVGL}), \textit{Group Fused Graphical Lasso} (GFGL; \cite{Gibberd.2017.GFGL}), \textit{Latent variable Time-varying Graphical Lasso} (LTGL; \cite{Tomasi.2018.LTGL}), \textit{Time Adaptive GGM} (TAGM; \cite{Tozzo.2021.TAGM}), \textit{Latent variable Time-varying Graphical Lasso with temporal kernel} (LTGL$_\kappa$; \cite{Tomasi.2021.TGL/LTG}), \textit{Latent variable Time-varying Graphical Lasso with Pattern inference} (LTGL$_p$; \cite{Tomasi.2021.TGL/LTG})). It seems worthwhile to assess in future work if these models can be applied to psychometric datasets, but this could encounter some challenges \parencite{LavelleHill.2025}.

\subsection{GGM --- "causal heuristics"}
When applying GGM's for causal discovery two heuristics pointed out in \textcite{Ryan.2022.SEM-DAG-GGM} might be useful. The first is that an edge between two variables in the undirected PMRF ($X_i-X_j$) indicates that two variables either share a direct causal link $\left(X_i \rightarrow X_j\right.$ or $\left.X_i \leftarrow X_j\right)$ or a common effect $\left(X_i \rightarrow X_k \leftarrow X_j\right)$. Furthermore, the absence of an edge between two variables indicates that these two variables do not share a direct causal link $\left(X_i \nrightarrow X_j\right.$ and $\left.X_i \nleftarrow X_j\right)$. Formally it can be stated that a undirected PMRF identifies the moral-equivalence set of the underlying DAG (i.e., collection of all DAG structures that produce the same moral graph, and hence the same PMRF), many different DAGs share the same moral graph \parencite{Ryan.2022.SEM-DAG-GGM}. GGMs therefore are a natural middle ground between zero-order correlations and DAGs and function as a bridge between the correlational and causal worlds \parencite{Epskamp.2017.LNM-RNM}.

\subsection{Random Fields (Statistics)}
The GGM and IM have been earlier already connected to the general concept of a Random Field \parencite{HernandezLemus.2021.MRF}, which is basically a joint probability distribution of a set of random variables, of which many variants with different degree of generality exist (see Figure \ref{fig:RandomFields}).
\begin{figure}[h]
    \centering
    \begin{tikzpicture}[
    node distance = 0.4cm and 0.4cm,
    every node/.style = {draw, rounded corners, minimum width=1cm, minimum height=0.5cm, font=\scriptsize\sffamily},
    every path/.style = {->, > = Stealth, thick}]
    \node (rf)         [fill=gray!20] {Random Field};
    \node (mrf)        [below=of rf] {MRF};
    \node (td-mrf)     [right=of mrf] {TD-MRF};
    \node (gmrf)       [below left=of mrf] {GMRF};
    \node (memrf)      [left =of mrf] {MEMRF};
    \node (hrf)        [above=of rf] {HRF};
    \node (d-hrf)      [right=of hrf] {D-HRF};
    \node (crf)        [left=of rf] {CRF};
    \draw (mrf)  -- (td-mrf);
    \draw (rf)   -- (mrf);
    \draw (rf)   -- (hrf);
    \draw (mrf)  -- (gmrf);
    \draw (mrf)  -- (memrf);
    \draw (hrf)  -- (d-hrf);
    \draw (rf)  -- (crf);
    \end{tikzpicture}
    \caption{Variants of Random Fields: \textit{Markov Random Field} (MRF; \cite{HernandezLemus.2021.MRF}), \textit{Gaussian MRF} (GMRF; \cite{Rue.2005.GMRF, Siden.2020.DeepGMRF}), \textit{Mixed Exponential MRF} (MEMRF; \cite{Yang.2014.MGM}), \textit{Time-Dynamic MRF} (TD-MRF; \cite{Cabello.2022.TD-MRF}), \textit{Conditional Random Field} (CRF; \cite{Sutton.2010.CRF}) \textit{Hybrid Random Field} (HRF; \cite{Kacprzyk.2011.HRF}), \textit{Dynamic Hybrid Random Field} (D-HRF; \cite{Bongini.2018.dHRF})}
    \label{fig:RandomFields}
\end{figure}
It seems interesting to investigate how Random Fields as the most basic statistical concept, relate to attempts of formalizing a general statistical model taxonomy in \parencite{Burkner.2023.Taxonomy} and \parencite[Chapter 12]{Getoor.2007.Taxonomy}. Also insightful to note here might be the literature on structure learning in Markov Random Fields \parencite{Drton.2017.StructureLearning, Mukherjee.2022.TensorIsing, Vogels.2024.StructureLearning}.

\subsection{Beyond Pairwise Correlations - Similarity \& Distance Measures}
Outside of psychometrics a large literature on a wide variety of statistical metrics is available as well. Similarity and Distance measures for Probability Density Functions, Vectors, and Graphs \parencite{Cha.2007, Josse.2016, Wills.2020, Das.2021, Subramanian.2022, Levy.2025}, also more specifically for categorical variables \parencite{vanVelden.2024}, and temporal graphs \parencite{Xiong.2026}, as well as for learning relations in time series data more generally \parencite{Fulcher.2013.hctsa, Fulcher.2017.hctsa, Lubba.2019.catch22}, and with different \textit{Pairwise Edge Measure} (PEMs) computed from lagged correlation matrices \parencite{Schwarze.2025.PEMs}. In \textcite{Wulkow.2022} a method for measuring dependencies between variables in a Dynamical System is proposed. Since some of this already intersects, it seems worthwhile to asses in more detail in the future if more of this is relevant and transferable in some form into psychometric applications.

\subsection{Dynamic Time Warping}
Additionally, some more novel approaches are available as well. \textit{Dynamic Time Warping} (DTW\footnote{[S1.2.7 Dynamic time warping (dtw)] in \parencite{Cliff.2023.SPI}}; \cite{tavenard.blog.DTW, Bringmann.2024.DTW, Kopland.2025.DTW}) has been developed in speech and signal processing for assessing the alignment between time series data and comes in many forms \parencite{Shifaz.2023.DTWvariants}. Which has recently been applied in psychology as well \parencite{Beurs.2024.DTW-applied, Pasteuning.2025.DTW-applied, Dingemans.2025.DTW-applied, Schwellnus.2025.DTW-applied}. It can complement highly parametrized approaches (e.g., mlVAR, DSEM) for exploratory identification of interesting patterns when usual assumptions are violated, for example stationarity \parencite{Ryan.2025.Stationarity}. DTW has been compared to mlVAR with strengths and weaknesses of both approaches assessed in more detail in \textcite{vanderDoes.2025.DTW-mlVAR}, and also combined with PGMs \parencite{Li.2018.DTW-GraphicalModels}.

\subsection{Attractor Landscapes}
A useful tool for investigating behavioral change in a complex dynamical system is the attractor\footnote{or Energy landscape \parencite{Masuda.2025.EnergyLands}, Potential/Stability landscape \parencite{Cui.2023.simlandr, Cui.2026.Isinglandr, Hoekstra.2026}} landscape \parencite{Heino.2023}. Along this line of thinking a method for analyzing ILD in the form of vector fields has been proposed with \textit{fitlandr} \parencite{Cui.2023.TV-VAR/fitlandr, fitlandr}, and potential landscapes in \textit{simlandr} \parencite{simlandr}.

\subsection{Recurrence Network}
Another based on Recurrence Plots (RPs; \url{http://www.recurrence-plot.tk/}) tool is the \textit{Recurrence Network} (RN\footnote{Correlation networks are a special case \parencite{Donner.2010.RN}, and Random Geometric Graphs (RGG; \cite{Duchemin.2023.RGG}) are formally equivalent \parencite{Zou.2019.RN}. Also related to Multiplex Graphs (or multilayer networks; \cite{Hasselman.2023}) which can be respectively constructed from Horizontal Visibility Graphs (HVG; \cite{Lacasa.2015.HVG}).}; \cite{Donner.2010.RN, Donner.2011.RN, Zou.2019.RN, Yang.2019.RN}). RNs provide additional quantitative metrics characterizing the dynamics in time-series data. Obtained by transforming the recurrence matrix of a RP into an adjacency matrix (unweighted and undirected network). 
\[
A_{t,t'}(\epsilon) = R_{t,t'}(\epsilon) - \delta_{t,t'} \tag{13} \label{eq_13}
\]
where $\delta_{t,t'}$ is the Kronecker delta to avoid artificial self-loops (a point is never connected to itself). Its topological characteristics preserve fundamental phase-space properties of the dynamical system \parencite{Donner.2010.RN}, and can represent different dynamical regimes \parencite{Donges.2015.pyunicorn}.

Common metrics are the \textit{Local} and \textit{Global Clustering Coefficient} which can provide information about the dimensionality of the underlying process \parencite{Varley.2021}, the \textit{Average Path Length} characterizes the average phase-space separation of states \parencite{Donner.2010.RN}, \textit{Transitivity} a measure of dynamical complexity \parencite{Zou.2019.RN}, and \textit{Assortativity} as a measure of the fragmentation of the attractor \parencite{Donner.2010.RN}. Since RNs and these graph-theoretical metrics are in their original application a discrete approximation of more general and continuous geometric properties of a dynamical system’s underlying attractor \parencite{Zou.2019.RN}, how this ultimately applies and transfers to psychometrics remains an open question as well.

Intuitively RNs can be understood as modeling the time evolution of a complex system as a trajectory through some kind of phase/state-space \parencite{Varley.2021}\footnote{points to a relation between RNs and \textit{Topological Data Analysis} (TDA; \cite{Chazal.2021.TDA, Sizemore.2019.TDA})}. And in some form this has already been applied in psychology previously \parencite{Coco.2014.RPinPsy, Wallot.2016.RPinPsy, Wallot.2018.RPinPsy, Wallot.2019.RPinPsy, Hasselman.2020.RN, Tomashin.2024.RPinPsy}, where the \textit{Cross-Recurrence Plot} (CRP) is a useful tool for analyzing the dynamics of dyads \parencite{Vanoncini.2025.CRQA, Logrieco.2026.CRQA, Renjaan.2026.CRQA}). \textit{Inter-system Recurrence Networks} (IRN) have been used to study the coupling dynamics between interacting systems of physiological variables \parencite{Hasselman.2023.RN}, and in \textcite{Hasselman.2022.EWS} a RN is used to analyze a depression-dataset (Idiographic; N=1, T=Large) for early warning signals \parencite{Evers.2024.EWS}.

\subsection{Higher-order Networks}
When Graph Theory is utilized to formulate our inferences problems, a large literature on higher order relations is---at least in principle---available for incorporation into future psychometric methodology as well. An introductory overview into the emerging field of networks beyond pairwise interactions like \textit{Hypergraphs} and \textit{Simplical-Complexes}\footnote{Have been identified as potentially appropriate for the social sciences some time ago already \parencite{Harary.1960.GraphTheroy-SocialScience}.} can be found in \textcite{Battiston.2020}. Hypergraphs have already been proposed in psychometrics to overcome the limitation of only modeling pairwise interactions \parencite{Marinazzo.2024.Hypergraphs}, and the IM can be formulated on hypergraphs as well \parencite{Mukherjee.2022.TensorIsing, Robiglio.2025.IsingHypergraph}, the SBM and ER model can be too \parencite{Mukherjee.2022.TensorIsing}. Related publications on hypergraphs include their application to Dynamical Systems and time-series data \parencite{Carletti.2020.Hypergraph, Delabays.2025.Hypergraph, Vaucher.2026.Hypergraphs} with community structure \parencite{Lotito.2024.Hypergraph, Ruggeri.2024.Hypergraph} and latent space approaches for graphs \parencite{Turnbull.2023.LatentSpace-Hypergraph}. When we attempt to model human subjective experiences---as we usually do in psychometrics---the variables of interest can originate from different levels of abstraction (e.g., items in psychometric tests referring to biological, psychological, or social variables), such a multi-level structure is also considered crucial in some theories of mind (e.g., Nested States Model (NSM; \cite{Denfield.2024.NSM})) and reflected in the application of \textit{multilayer networks} \parencite{Kivela.2014.Multilayer} as a framework for more adequately modeling psychopathologies \parencite{Boer.2021}, as well as for ecological modeling in general \parencite{Jordan.2022}, and represents yet another important methodological extension \parencite{Kim.2020, Krishnagopal.2023}. In addition to all that, there is a huge literature on investigating higher-order interactions in dynamical systems and how important they are for understanding their dynamics \parencite{Lambiotte.2019.HigherOrder, Klein.2020.HigherOrder, Battiston.2021.HigherOrder, Yao.2021.HigherOrder, Majhi.2022.HigherOrder, Rosas.2022.HigherOrder, Bick.2023.HigherOrder, Boccaletti.2023.HigherOrder, Ceria.2023.HigherOrder, Malizia.2024.HigherOrder, Muolo.2024.HigherOrder, Malizia.2025.HigherOrder, Papillon.2025.HigherOrder, Robiglio.2025.HigherOrder}.

\printbibliography
\end{document}